\documentclass[12pt]{article}
\usepackage[english]{babel}
\usepackage[normalem]{ulem}
\usepackage{multirow}
\usepackage{booktabs}
\usepackage{float}
\usepackage[text={15.5cm,22.5cm}]{geometry}
\usepackage{graphicx}
\usepackage[dvipsnames]{xcolor}
\usepackage{pdfcolmk}
\usepackage{amsmath}
\usepackage{amssymb}
\usepackage{slashed}
\usepackage{hyperref}
\usepackage{tabularx}
\usepackage{cite}
\allowdisplaybreaks

\newcommand{\DM}{\mathrm{DM}}

\newcommand{\GeV}{~\mathrm{GeV}}
\newcommand{\TeV}{~\mathrm{TeV}}

\begin{document}

\title{\vspace{-0.8cm}
{\normalsize
\flushright TUM-HEP 1066/16\\ LMU-ASC 53/16\\MPP-2016-320\\}
\vspace{1cm}
\bf Dark matter decay through gravity portals
 \\ [8mm]}

\author{Oscar Cat\`a$^1$, Alejandro Ibarra$^2$, Sebastian Ingenh\"utt$^{2,3}$\\[2mm]
{\normalsize\it $^1$ Ludwig-Maximilians-Universit\"at M\"unchen, Fakult\"at f\"ur Physik,}\\[-0.05cm]
{\normalsize\it Arnold Sommerfeld Center for Theoretical Physics, 80333 M\"unchen, Germany}\\[2mm]
{\normalsize\it $^2$ Physik-Department T30d, Technische Universit\"at M\"unchen,}\\[-0.05cm]
{\it\normalsize James-Franck-Stra\ss{}e, 85748 Garching, Germany}\\[2mm]
{\normalsize \it $^3$ Max-Planck-Institut f\"ur Physik (Werner-Heisenberg-Institut),}\\ [-0.05cm]
{\normalsize \it F\"ohringer Ring 6, 80805 M\"unchen, Germany}
}

\date{}
\maketitle
\thispagestyle{empty}
\vskip 1.5cm
\begin{abstract}
Motivated by the fact that, so far, the whole body of evidence for dark matter is of gravitational origin, we study the decays of dark matter into Standard Model particles mediated by gravity portals, i.e., through nonminimal gravitational interactions of dark matter. We investigate the decays in several widely studied frameworks of scalar and fermionic dark matter where the dark matter is stabilized in flat spacetime via global symmetries. We find that the constraints on the scalar singlet dark matter candidate are remarkably strong and exclude large regions of the parameter space, suggesting that an additional stabilizing symmetry should be in place. In contrast, the scalar doublet and the fermionic singlet candidates are naturally protected against too fast decays by gauge and Lorentz symmetry, respectively. For a nonminimal coupling parameter $\xi\sim {\cal{O}}(1)$, decays through the gravity portal are consistent with observations if the dark matter mass is smaller than $\sim 10^5$ GeV, for the scalar doublet, and $\sim 10^6$ GeV, for the fermionic singlet.
\end{abstract}
\newpage
\section{Introduction}
\label{sec:intro}

A myriad of astronomical and cosmological measurements have proven to be inconsistent with the predictions of General Relativity, should the matter content of galaxies, clusters of galaxies and of the large-scale Universe be that of the Standard Model (SM). A plausible explanation for this long-standing puzzle consists in postulating the existence of a non-luminous matter component in our Universe, dubbed dark matter (DM), possibly requiring physics beyond the Standard Model (for reviews, see ~\cite{Bertone:2010zza,Bergstrom:2000pn,Bertone:2004pz}). Unfortunately, despite being responsible for the largest fraction of the matter content in our Universe, very little is known about the nature or the interactions of dark matter.

One of the most intriguing aspects of dark matter is its large lifetime, which exceeds the age of the Universe, since it was already present at the time of Big Bang Nucleosynthesis and has survived until present times. Furthermore, if dark matter decays producing photons, neutrinos or antimatter, the requirement that the cosmic fluxes of these particles do not exceed observations translates into limits on the lifetime which can be orders of magnitude larger than the age of the universe (for a review, see~\cite{Ibarra:2013cra}).

Particles can be long-lived due to kinematical reasons, dynamical reasons (in the form of a protecting symmetry) or a combination of both. Among the vast list of known particles, only very few have lifetimes longer than the age of the Universe. The lightest neutrino, the electron and the proton are long-lived due to the conservation of the Lorentz symmetry, the electric charge and the total baryon number, respectively, while the two other neutrino mass eigenstates are long-lived as a consequence of the small neutrino mass splittings, which translates into a tiny phase space available in their decay. For dark matter, on the other hand, the mechanism responsible for its longevity is not yet known. 

For very light dark matter, various plausible scenarios have been proposed ensuring a sufficiently long lifetime based on kinematical arguments; a paradigmatic example is the axion~\cite{Abbott:1982af,Preskill:1982cy,Dine:1982ah}, which is predicted to have a mass of a few $\mu$eV and decays into two photons with a lifetime of $\sim 10^{54}\,{\rm s}$. However, for most dark matter candidates, the phase space available in the decay is sizable and therefore a dynamical mechanism must be at work in order to reduce the rate, usually in the form of a small coupling between the parent and the daughter particles, which may be attributed to a mildly broken global or continuous symmetry. A renowned example is the keV-scale sterile neutrino~\cite{Dodelson:1993je,Shi:1998km,Abazajian:2001nj} which is a phenomenologically viable dark matter candidate only when the mixing angle with the active neutrino is very small, e.g., for $m_{\rm DM} \simeq 5$ keV the mixing angle must be $\sin^2(2\theta)\sim 10^{-9}-10^{-12}$.

For dark matter produced via thermal freeze-out, which favors masses between 100 MeV~\cite{Boehm:2003hm} and 100 TeV~\cite{Griest:1989wd}, or for non-thermally produced dark matter, for which the mass can be as large as the Grand Unification scale \cite{Chung:1998ua}, the stabilization problem becomes even more acute. For instance, the scalar singlet candidate, $\phi$, could decay into a pair of Higgs bosons, $h$, via the Lagrangian term $\mu \phi hh$, with $\mu$ a mass parameter. Then, the requirement that the dark matter has a lifetime longer than the age of the Universe translates, for $m_{\rm \phi}=500$ GeV, into $\mu\lesssim 3\times 10^{-19}$ GeV.  A common, and admittedly ad hoc, justification for the smallness of the parameters inducing dark matter decay consists in postulating a global symmetry that precludes the destabilizing couplings altogether.

This stabilizing symmetry is normally imposed on the flat spacetime Lagrangian. A relevant question, which is usually ignored, is whether the stabilizing mechanism remains operational in the presence of spacetime curvature effects. After all, all current evidence for dark matter so far comes from its gravitational interactions with ordinary matter. In this work we will explore the possibility that dark matter could decay via interactions that vanish in the absence of gravitational interactions but are present in curved spacetime, which we denominate the ``gravity portal''. From the possible interaction operators, we will restrict our attention to those with the lowest dimensionality, which are expected to yield the dominant effects. With this requirement, the dominant operator turns out to be unique and proportional to the Ricci scalar $R$. If present in the Lagrangian, this term will lead to dark matter decay with a rate suppressed by inverse powers of the Planck mass, giving rise to a long, albeit potentially measurable, lifetime. 

In a recent paper~\cite{Cata:2016dsg} we have explored such gravity portals for the simplest dark matter candidate, namely the scalar singlet. Here we will extend the analysis to other dark matter candidates commonly considered in the literature: the scalar inert doublet and the fermionic singlet. Our focus will be on the constraints on the decay width from gravity portal interactions and we will not address other aspects of dark matter phenomenology, such as production in the early Universe or the scattering rate with nuclei, which are not necessarily linked to the gravity portal.

This paper will be organized as follows: in Section 2 we will introduce the general setup and discuss the generic properties expected from dark matter decaying through gravity portals. This will be illustrated with specific examples in Section 3, where we concentrate on scalar singlet, scalar inert doublet and fermionic singlet dark matter realizations. We investigate the different decay channels into Standard Model particles for each of them and place the corresponding constraints on the gravitational interactions. We summarize in Section 4. Technical details are collected in the appendices.  

\section{Dark matter decay via gravity portals}
\label{sec:setup}
We consider the Standard Model, with fields denoted as $X$, extended by a dark matter field, denoted by $\varphi$ (Lorentz or spinorial indices are omitted but should be implicitly understood). In the absence of gravity, the action can be cast as:
\begin{equation} \label{eq:CAJ}
\mathcal S = \int d^4x \left[\mathcal L_{\rm SM}(X) + \mathcal L_{\DM}(\varphi,X)\right]\;,
\end{equation}
where $\mathcal L_{\rm SM}(X)$ is the Standard Model Lagrangian and $\mathcal L_{\DM}(\varphi,X)$ contains terms involving just the dark matter field as well as possible interaction terms with the Standard Model fields;  the concrete form of these interactions is however inessential for our discussion. We also assume, as is commonly done in the literature, the existence of a stabilizing global symmetry, under which the Standard Model fields, but not the dark matter field, transform trivially. 

In the presence of gravitational interactions, the stabilizing symmetry remains unbroken provided the dark matter only couples minimally to gravity. Nevertheless, nonminimal coupling to gravity may break the global symmetry and therefore induce dark matter decay. The dominant operators will be the ones with the lowest dimension. It can be checked that, regardless of the specific dark matter candidate $\varphi$, the most relevant operators will be those proportional to the Ricci scalar $R$ times a function linear in $\varphi$.\footnote{Higher powers of $R$ will bear further suppressions in powers of $M_P^{-1}$, while terms proportional to $R_{\mu\nu}$,  e.g. $R^{\mu\nu}\partial_{\mu}\varphi\partial_{\nu}\varphi$ are at least of dimension 6.} The action then takes the form
\begin{equation} \label{eq:CompleteActionJordan}
  \mathcal S = \int d^4x \sqrt{-g}\left[ -\frac{R }{2 \kappa^2}+ \mathcal L_{\rm SM}(X) + \mathcal L_{\DM}(\varphi,X) -\xi R f(\varphi,X)\right]\;,
\end{equation}
where $g$ is the determinant of the metric tensor $g_{\mu\nu}$, $\kappa=\overline M_P^{-1}=\sqrt{8 \pi G}$ is the inverse (reduced) Planck mass and $\xi$ is a dimensionless coupling. $\mathcal L_{\rm SM}$ is the Standard Model Lagrangian in a general gravitational background, which can be cast as
\begin{align}
{\mathcal{L}}_{\rm SM}&\equiv {\mathcal{T}}_{F}+{\mathcal{T}}_{f}+{\mathcal{T}}_{H}+{\mathcal{L}}_{Y}-V_{H}\;,
\end{align}
where ${\cal{L}}_Y$ contains the Yukawa interactions, $V_{H}$ is the Higgs potential and ${\mathcal{T}}_j$ are the kinetic terms of gauge bosons, fermions and scalars:
\begin{align}
{\mathcal{T}}_{F}&=-\frac{1}{4}g^{\mu\nu}g^{\lambda\rho}F_{\mu\lambda}^aF_{\nu\rho}^a\;,\nonumber\\
{\mathcal{T}}_{f}&=\frac{i}{2}{\overline{f}}\stackrel{\longleftrightarrow}{\slash{\!\!\!\!\nabla}}\!\!f\;,\nonumber\\
{\mathcal{T}}_{H}&=g^{\mu\nu}(D_{\mu}H)^{\dagger}(D_{\nu}H)\;.
\end{align}
Above we have defined $\slash{\!\!\!\!\nabla}=\gamma^ae^{\mu}_a\nabla_{\mu}$, $\gamma^a$ being a Dirac matrix, $e^{\mu}_a$ a vierbein and ${\nabla}_\mu=D_\mu -\frac{i}{4}e_\nu^b (\partial_\mu e^{\nu c}) \sigma_{bc}$, with $D_\mu$ the gauge covariant derivative.

The action in Eq.~(\ref{eq:CompleteActionJordan}) is written in the so-called Jordan frame. In this frame the equations of motion for $\varphi$ and the metric tensor are coupled through the nonminimal operators. In order to select a well-defined gravitational background (Minkowski in our case), it is convenient to make a field redefinition to get the two fields decoupled. This is the so-called Einstein frame, where the equations of motion for the gravitational part are simply Einstein's equations, which are easier to solve than the coupled ones that result in the Jordan frame. 

Given the class of nonminimal operators that we are considering, the transition to the Einstein frame is operated through a  Weyl transformation on the metric tensor. More specifically, the action in the Jordan frame can be recast as
\begin{equation} \label{eq:CompleteActionJordanDM}
  \mathcal S = \int d^4x \sqrt{-g}\left[ -\frac{R }{2 \kappa^2}\Omega^2(\varphi,X) + \mathcal L_{\rm SM} + \mathcal L_{\DM} \right]\;,
\end{equation}
where 
\begin{equation}\label{eq:def-Omega}
\Omega^2(\varphi,X) =1 + 2 \kappa^2 \xi f(\varphi,X)\;.
\end{equation}
The dark matter field and the Ricci scalar can be decoupled by redefining the metric tensor as
\begin{equation}\label{eq:transfo}
  \widehat g_{\mu\nu} = \Omega^2(\varphi,X) g_{\mu\nu}\;,
\end{equation} 
which in turn implies the following relation between the Ricci scalar in the Einstein and Jordan frames~\cite{Salopek:1988qh}:
\begin{align}
\widehat R=\frac{1}{\Omega^{2}}R-\frac{6}{\Omega^{3}} g_{\mu\nu}\nabla^\mu\nabla^\nu \Omega\;.
\end{align}
Finally, the action expressed in the Einstein frame reads:
\begin{equation} \label{eq:CompleteActionEinstein}
  \mathcal S=\int d^4x \sqrt{- \widehat g}\left[ -\frac{\widehat R}{2 \kappa^2} + \frac{3}{\kappa^2}\widehat{g}_{\mu\nu}\frac{\widehat \nabla^\mu \Omega \widehat \nabla^\nu \Omega}{\Omega^2} + \mathcal {\widehat{L}}_{\rm SM} +\mathcal {\widehat{L}}_{\DM} \right]\;,
\end{equation}
where all hatted quantities are constructed in terms of the Weyl-transformed metric tensor $\widehat g_{\mu\nu}$.\footnote{The transition between both frames is operated through a field redefinition, and therefore they are equivalent, regardless of whether one treats the gravitational field as a classical background or as a quantum one. For instance, at the quantum level (within linearized gravity) nonminimal operators linear in $\varphi$ and $R$ potentially generate kinetic mixing between dark matter and the graviton. In this limit, the transition to the Einstein frame can be seen as equivalent to the diagonalization of the kinetic terms~\cite{Ren:2014sya}.} In particular, $\mathcal {\widehat{L}}_{\rm SM}$ and  $\mathcal {\widehat{L}}_{\DM}$ are given by
\begin{align} \label{eq:SMLagrangianEinstein}
{\mathcal{\widehat{L}}}_{\rm SM}&={\mathcal{\widehat T}}_{F}+\frac{1}{\Omega^{3}}{\mathcal{\widehat T}}_{f}+\frac{1}{\Omega^2}{\mathcal{\widehat T}}_{H}+\frac{1}{\Omega^4}({\mathcal{L}}_{Y}-V_{H}) 
\end{align}
and
\begin{align}
\mathcal {\widehat{L}}_{\DM}=\frac{1}{\Omega^{n}}{\mathcal{\widehat T}}_{\varphi}-\frac{1}{\Omega^{4}}V(\varphi,X)\;,
\end{align}
where $n=2$ or $3$ depending, respectively, on whether dark matter is bosonic or fermionic. 

Upon Taylor expanding $\Omega(\varphi,X)$ in $\varphi$ one finds interaction terms in the Lagrangian that lead to dark matter decay into Standard Model particles. Explicitly, the first-order term in the expansion of the Weyl-transformed Standard Model Lagrangian reads:
\begin{align} \label{eq:SMLagElinear}
{\mathcal{\widehat{L}}}_{{\rm SM},\varphi}&=-2\kappa^2\xi  \left.\frac{\partial f}{\partial \varphi}\right|_{\varphi=0} \varphi\left[\frac{3}{2}{\mathcal{\widehat T}}_{f}+{\mathcal{\widehat T}}_{H}+2({\mathcal{L}}_{Y}-V_{H})\right]\;.
\end{align}
The first-order term in the expansion of ${\cal{\widehat L}}_{\rm DM}$ is cubic in $\varphi$ and only induces dark matter decay after closing a loop with two $\varphi$-legs; this term will be neglected in our analysis, along with all higher order terms in the expansion. On the other hand, the term in Eq.~(\ref{eq:CompleteActionEinstein}) of the form $\Omega^{-2} \nabla^\mu \Omega \nabla^\nu \Omega$ may lead, depending on the form of $f(\varphi,X)$, to an additional contribution to the dark matter kinetic Lagrangian after Taylor expanding. The kinetic term must then be brought back into the canonical form via a field rescaling. However, the correction to the field is ${\cal{O}}(\xi^2)$ and, in view of the constraints on $\xi$ to be found later in this paper, can be safely neglected. Similarly, for scalar dark matter scenarios the gravity portal shifts the vacuum away from $\langle H\rangle=v/\sqrt{2}$, $\langle \varphi\rangle=0$ with corrections of ${\cal{O}}(\xi)$. However, these effects show up at the amplitude level at least at ${\cal{O}}(\xi^2)$ and can be consistently dismissed. Therefore, the phenomenology of the dark matter decay via the gravity portal can be well described by considering only the Lagrangian of Eq.~(\ref{eq:SMLagElinear}). 

Given the structure of Eqs.~(\ref{eq:SMLagrangianEinstein}) and (\ref{eq:SMLagElinear}), there are a number of generic aspects of this mechanism of dark matter decay that hold independently of the specific form of the dark matter candidate:
\begin{itemize}
\item The nonminimal coupling, when expressed in the Einstein frame, leads to an infinite number of operators of the form
\begin{align}
\left(\xi\frac{\varphi}{\overline{M}_P^2}\right)^n\left[\frac{3}{2}{\mathcal{\widehat T}}_{f}+{\mathcal{\widehat T}}_{H}+2({\mathcal{L}}_{Y}-V_{H})\right]\;,
\end{align}
which are suppressed by powers of the nonminimal coupling parameter $\xi$ and the inverse of the Planck mass. Therefore, the branching ratios (BRs) of processes involving $n$ dark matter particles, and in particular the decay branching ratios (corresponding to $n=1$),  are determined by the Weyl weights and known Standard Model parameters, the only free parameter being the dark matter mass.
\item The full set of higher dimensional operators that results from expanding the Weyl factor does not originate from integrating out heavy particles, as is the case in weakly-coupled new physics extensions. Instead, these operators arise as a consequence of the field redefinition of Eq.~(\ref{eq:transfo}).
\item  The Taylor expansion not only generates operators linear in $\varphi$, which induce dark matter decay, but also operators quadratic in $\varphi$, which lead to dark matter annihilation. The latter are further suppressed by additional powers of $M_P$ and do not contribute significantly to the total annihilation rate nor to the dark matter freeze-out, unless $\xi\gg 1$~\cite{Ren:2014mta}. Such large values of the nonminimal coupling parameter are, however,  at odds with observations for the range of dark matter masses favored by thermal production, as we will show in this paper.
\item Since the gauge kinetic term remains invariant under the Weyl transformation, there is no local coupling of dark matter to massless gauge bosons. However, since the Standard Model is not conformal invariant, decay into pairs of massless gauge bosons do occur at the loop level.
\item Couplings to the $Z$ and the $W$ gauge bosons are induced through the kinetic term of the Higgs boson. Dark matter therefore couples to the longitudinal polarizations. By the Goldstone equivalence theorem, this implies that the decays into $Z$ and $W$ will tend, up to phase space suppressions, to dominate the decay width. 
\item The cosmological constant induces dark matter decay into gravitons through an operator of the form $\xi\kappa^{2}\varphi \sqrt{-g}\Lambda$.  However, in view of the tiny value of the cosmological constant which follows from observations, $\Lambda\sim (10^{-3}\,{\rm eV})^4$, this contribution will be extremely suppressed compared to the one induced by Standard Model loops.\footnote{Dark matter decays into gravitons can be relevant in some other scenarios, such as the one considered in \cite{Tang:2016vch}, where the global symmetry is broken by terms already present in the flat spacetime action.}
\end{itemize}

We remark that the gravitational portal is a specific realization of an effective field theory for dark matter decay, where the vertex topologies are those of the Standard Model with one extra $\varphi$ leg. As such, one can distinguish the topologies related to the Higgs sector, which are proportional to the masses of the Standard Model particles, from the ones related to the gauge interaction sector, which are mass-independent. More specifically, in the gravity portal framework, the Wilson coefficients associated with the effective operators $\varphi V^\mu V_\mu h^n$, $\varphi \overline{f} f  h^n$ and $\varphi h^n$ are proportional to $m_V^2$, $m_f$ and $F_n(m_\phi^2, m_h^2)$, respectively, where $n=0,1,2$ labels the number of Higgs legs in the vertex and $F_n$ is a function that depends on $n$. On the other hand, the Wilson coefficient for $\varphi{\overline{f}}\gamma_{\mu}f V^{\mu}h^n$ depends on the gauge coupling of the gauge boson $V^\mu$ to the fermion $f$, but is independent of their masses. Operators of the form $\varphi V^{\mu\nu}V_{\mu\nu}$, with $V^{\mu\nu}$ a field strength tensor, are absent at tree level; decays into pairs of massless gauge bosons can therefore only take place at loop level.

Note that our approach predicts a generic suppression of $1/M_P^2$ for each field $\varphi$, regardless of the dimensionality of the operator.\footnote{This is at variance with other phenomenological analyses of dark matter decay via Planck mass suppressed operators; see e.g.~\cite{Mambrini:2015sia}.} Therefore, increasing the number of $\varphi$ fields on each of the operators above (which can easily be done by expanding the Weyl function $\Omega$ beyond linear order in $\varphi$) provides extra factors of $1/M_P^2$ but does not modify the generic Standard Model parameter dependence of the operators. The exact analytical form of the resulting vertices depends on the specific dark matter candidate. The relevant vertices for decay can be found in Appendix \ref{app:rules}.  

\section{Constraints on specific dark matter scenarios}
\label{sec:pheno}

In this section we particularize the mechanism of dark matter decay via gravity portals to some simple frameworks, and we calculate the predicted branching ratios for various decay modes, as well as the constraints on the model parameters stemming from observations. We will first revisit the scalar singlet dark matter scenario, and we will expand the analysis and the discussion presented in~\cite{Cata:2016dsg}. We will then extend the analysis to the dark matter scalar doublet and fermionic singlet dark matter scenarios. In this paper we will restrict our analysis to dark matter masses above the QCD confinement scale, such that the phenomenology can be appropriately described using the Standard Model degrees of freedom. For dark matter masses below the confinement scale different methods have to be used; we plan to study gravitational portals for light dark matter in a forthcoming paper~\cite{CII:2016}.

In order to constrain the strength of the gravity portal one can impose the necessary condition that the dark matter lifetime is longer than the age of the Universe, $\tau_U\sim 4\times 10^{17}$ s. However, stronger limits follow from the requirement that the fluxes of gamma-rays, antimatter and neutrinos predicted to be produced in the decay do not exceed observations. These limits have been investigated in detail for two-body decay channels, such as $WW,hh,{\overline{f}}f$. For instance, for decays into $WW$, the inverse width must be $\Gamma_{WW}^{-1}\gtrsim 10^{27}$ s for $m_{\rm DM}\sim 200$ GeV - $30$ TeV from measurements of the gamma-ray flux with the Fermi-LAT~\cite{Cirelli:2012ut}. Competitive results are also found from the positron data~\cite{Ibarra:2013zia} and antiproton data~\cite{Giesen:2015ufa} measured by AMS-02~\cite{Accardo:2014lma,Aguilar:2016kjl}. In the high-mass regime, neutrino telescopes set the most stringent bounds; for decays into $\nu\overline \nu$, the inverse width must be $\Gamma_{\nu\overline\nu}^{-1}\gtrsim 10^{26}$ s for $m_{\rm DM}\sim 10$ TeV - $10^{15}$ GeV~\cite{Esmaili:2012us}. Decays with more particles in the final state, on the other hand, are less studied. In particular, the four-, five- and six-body decay channels that, as will see below, dominate the decay for large dark matter masses have never been considered in the literature. A detailed study of the exclusive decay channels is out of the scope of this work. Therefore, we will conservatively adopt a mass-independent lower bound on  the total inverse width, $\Gamma^{-1}_{\rm tot}\gtrsim 10^{24}\,{\rm s}$.

\subsection{Scalar singlet dark matter}
\label{sec:scalar}
Consider a real scalar field $\phi$, singlet under the Standard Model gauge group and charged under a discrete $Z_2$ symmetry under which $\phi$ is odd ($\phi \to -\phi$) whereas all Standard Model fields are even~\cite{Silveira:1985rk,McDonald:1993ex,Cline:2013gha}. For this particular case, the nonminimal coupling term is of canonical dimension 3, and the full theory reads, in the Jordan frame,  
\begin{equation} \label{eq:CompleteActionJordanphi}
  \mathcal S = \int d^4x \sqrt{-g}\left[ -\frac{R }{2 \kappa^2}+ \mathcal L_{\rm SM}(X) + \mathcal L_{\DM}(\phi,H) -\xi M R \phi\right]\;,
\end{equation}
where $M$ is a mass scale, possibly close to $M_P$. Keeping only terms up to dimension four, ${\cal{L}}_{\DM}$ is given by
\begin{equation}
  \mathcal L_{\DM}=\frac{1}{2}g_{\mu\nu}\partial^\mu \phi \partial^\nu \phi -V(\phi,H)\;,
\end{equation}
where the interaction terms in the potential are assumed to be invariant under the transformation $\phi\rightarrow -\phi$. The most general potential then takes the form
\begin{align}
V(\phi,H)=\frac{1}{2}\mu^2_{\phi}\phi^2+\frac{1}{4!}\lambda_{\phi}\phi^4+\frac{1}{2}\lambda_{\phi H}\phi^2(H^{\dagger}H)\;,
\end{align}
where the last term describes the Higgs portal interaction, which may be responsible, for appropriate values of the coupling strength $\lambda_{\phi H}$, for generating the observed dark matter density via thermal freeze-out.

We now express the Lagrangian in the Einstein frame via the field redefinition Eq.~(\ref{eq:transfo}) with
\begin{align}
\Omega^2(\phi)=1+2\xi M\kappa^2\phi\;,
\end{align}
and Taylor expand the Weyl-transformed Lagrangian in powers of $\phi$. As emphasized in the previous section, only ${\cal{\widehat L}}_{\rm SM}$ contains linear terms in $\phi$ which induce tree-level decay. Explicitly,
\begin{align} \label{eq:LSM_ScalarDM_Einstein}
{\mathcal{\widehat{L}}}_{{\rm SM},\phi}&= -2\xi M\kappa^2\phi\left[\frac{3}{2}{\mathcal{\widehat T}}_{f}+{\mathcal{\widehat T}}_{H}+2({\mathcal{L}}_{Y}-V_{H}) \right]\;.
\end{align}

The full set of effective decay vertices generated from Eq.~(\ref{eq:LSM_ScalarDM_Einstein}) is listed in Appendix \ref{app:rules}. Note that the specific form of the dark matter interactions implies that up to four-body decays are possible at tree level: the two-body decays arise from the kinetic and mass terms of the Standard Model particles, while three- and four-body decays arise from the interaction operators in the Standard Model, augmented by a single power of $\phi$. In Table~\ref{tab:phiDecays} we list all possible tree-level decay channels, together with the kinematic dependence of the rates, omitting the common prefactor $\xi^2 M^2\kappa^4$ and the phase space factors.  
Decays involving $W$ and $Z$ gauge bosons dominate whenever allowed by phase space. The dominant mode is actually the one with longitudinal components, as follows from the Goldstone equivalence theorem. This explains the relative $m_{\phi}^2/v^2$ enhancement of $\phi\to Z(W){\overline{f}}f^{(\prime)}$ with respect to $\phi\to f\overline{f} (g,\gamma)$. The $m_{\phi}^2/v^2$ enhancement factors in $\phi\to WW h$ and $\phi\to WWhh$ with respect to $\phi\to WW$ simply compensate dimensionally for each added Higgs field. For decay modes with the same scaling, e.g. $\phi\to WWh$ vs. $\phi\to f\overline{f}Z$, phase space considerations determine which are the dominant ones. 

The exact analytical expressions for the rates of the two-body decay channels read:
\begin{align}
\Gamma_{\phi\to hh}&=\frac{\xi^2M^2\kappa^4}{32\pi}m_{\phi}^3(1+2x_h)^2(1-4x_h)^{1/2}\;,\nonumber\\
\Gamma_{\phi\to ZZ}&=\frac{\xi^2M^2\kappa^4}{32\pi}m_{\phi}^3(1-4x_Z+12x_Z^2)(1-4x_Z)^{1/2}\;,\nonumber\\
\Gamma_{\phi\to WW}&=\frac{\xi^2M^2\kappa^4}{16\pi}m_{\phi}^3(1-4x_W+12x_W^2)(1-4x_W)^{1/2}\;,\nonumber\\
 \Gamma_{\phi\to \overline{f}_{\!i} f_i}&=N_c^{(f_i)}\frac{\xi^2M^2\kappa^4}{8\pi}m_{\phi}^3\,x_{\! f_i}(1-4x_{\!f_{i}})^{3/2}\;,
\end{align} 
where $i$ is a flavor index and $x_a=m_a^2/m_{\phi}^2$. For three- and four-body decay channels, the decay rates can be obtained following the procedure described in Appendix \ref{app:kinematics}, where we discuss how to efficiently account for the kinematics of $n$-body decays.  The exact expressions are rather cumbersome, but simpler approximate expressions can be obtained by taking the limit $m_{\phi}\gg m_{\rm SM}$, with $m_{\rm SM}$ the largest mass of the Standard Model particles in the final state. The results are:
\begin{align}
\Gamma_{\phi\to \overline{q}_i q_i g}&\simeq \alpha_s\frac{\xi^2M^2\kappa^4}{4\pi^2}m_{\phi}^3\;,\nonumber\\
\Gamma_{\phi\to {\overline{f}_{\!i}}f_j^\prime W}&\simeq \frac{3}{4\sqrt{2}}G_FN_c^{(f_i)}|U_{ij}|^2\frac{\xi^2M^2\kappa^4}{(4\pi)^3}m_{\phi}^5\;,\nonumber\\
\Gamma_{\phi\to {\overline{f}_{\!i}}f_iZ}&\simeq \frac{3}{2\sqrt{2}}G_FN_c^{(f_i)}(g_V^2+g_A^2)\frac{\xi^2M^2\kappa^4}{(4\pi)^3}m_{\phi}^5\;,\nonumber\\
\Gamma_{\phi\to WWhh}&\simeq\frac{\xi^2M^2\kappa^4}{15(8\pi)^5v^4}m_{\phi}^7\;,\nonumber\\
\Gamma_{\phi\to ZZhh}&\simeq\frac{\xi^2M^2\kappa^4}{30(8\pi)^5v^4}m_{\phi}^7\;,
\end{align} 
where $g_V=\frac{1}{2}t_3-Q\sin^2\theta_W$, $g_A=\frac{1}{2}t_3$ and $U_{ij}$ is a matrix relating the weak eigenstates to the mass eigenstates. For quarks, $U$ is the CKM matrix; for leptons, the PMNS matrix. Here and henceforth, rates for processes denoted with a single $W$ in the final state should be understood as the rate for the process plus its complex conjugate, for example, $\Gamma(\phi\to {\overline{f}_{\!i}}f_j^\prime W)\equiv \Gamma(\phi\to {\overline{f}_{\!i}}f_j^\prime W^+)+\Gamma(\phi\to \overline{f^\prime }_{\!\!\!j} f_i W^-)$.

\begin{table}[t!]
\centering
\renewcommand{\arraystretch}{1.2}
  \begin{tabular}{| l | c |}
    \hline
    \hskip 1.0cm Decay mode &  Scaling \\
    \hline
    $\phi\rightarrow hh, WW, ZZ$                          & $m_\phi^3$ \\
    $\phi\rightarrow f\overline f$		                        & $m_f^2 m_\phi$\\
    \hline
    $\phi\rightarrow hhh$			                        & $v^2m_\phi$\\
    $\phi\rightarrow WWh, ZZh$	                        & $m_\phi^5/v^2$\\
    $\phi\rightarrow f\overline fh$		                        & $m_f^2 m_\phi^3/v^2$\\
    $\phi\rightarrow f\overline f'W, f\overline fZ$	& $m_\phi^5/v^2$\\
    $\phi\rightarrow f\overline f\gamma, q\overline qg$	& $m_\phi^3$\\
    \hline
    $\phi\rightarrow hhhh$			                        & $m_\phi^3$\\
    $\phi\rightarrow WWhh, ZZhh$			& $m_\phi^7/v^4$\\
    \hline
  \end{tabular}
  \caption{Tree-level decay channels of the scalar singlet dark matter candidate induced by the nonminimal coupling to gravity, together with the parametric dependence on the dark matter mass and the Standard Model mass scales.}
  \label{tab:phiDecays}
\end{table}

\begin{figure}[t!]
\begin{center}
 \includegraphics[width=0.49\textwidth]{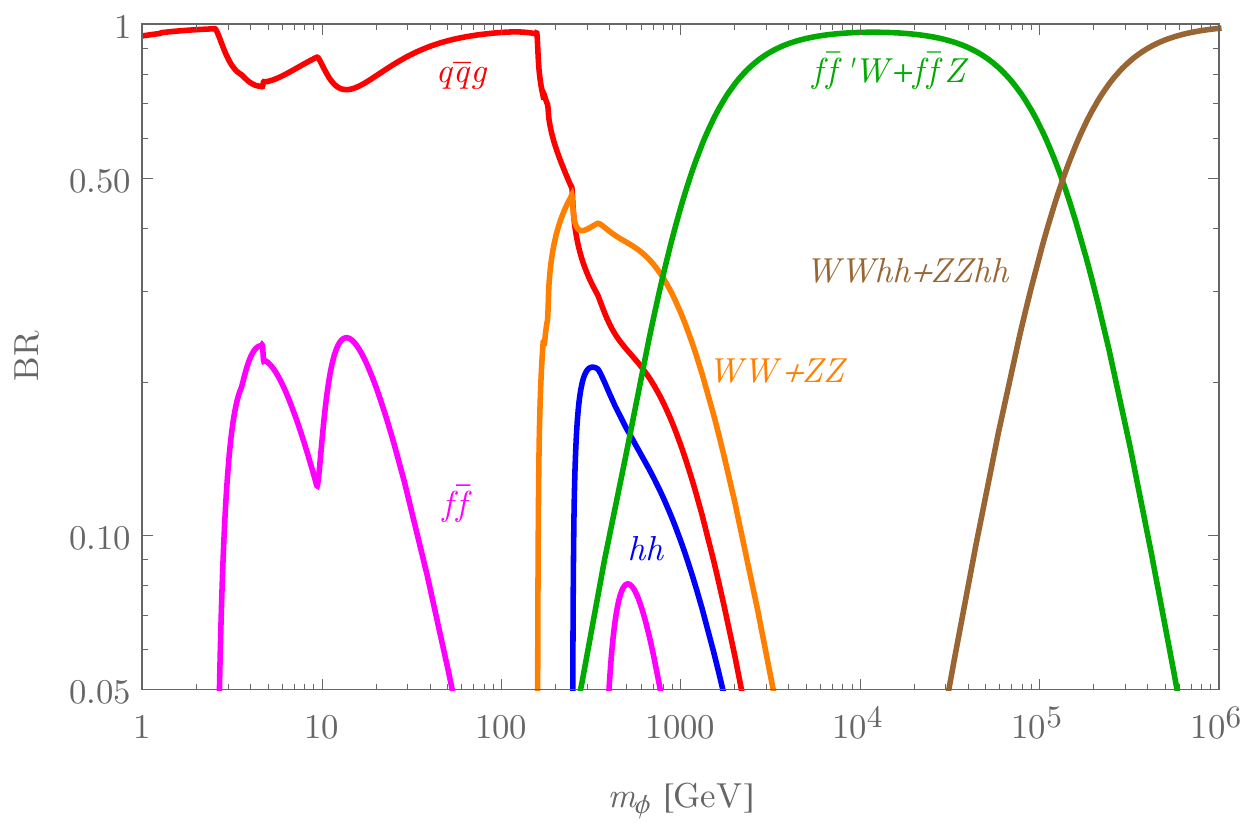} 
 \includegraphics[width=0.49\textwidth]{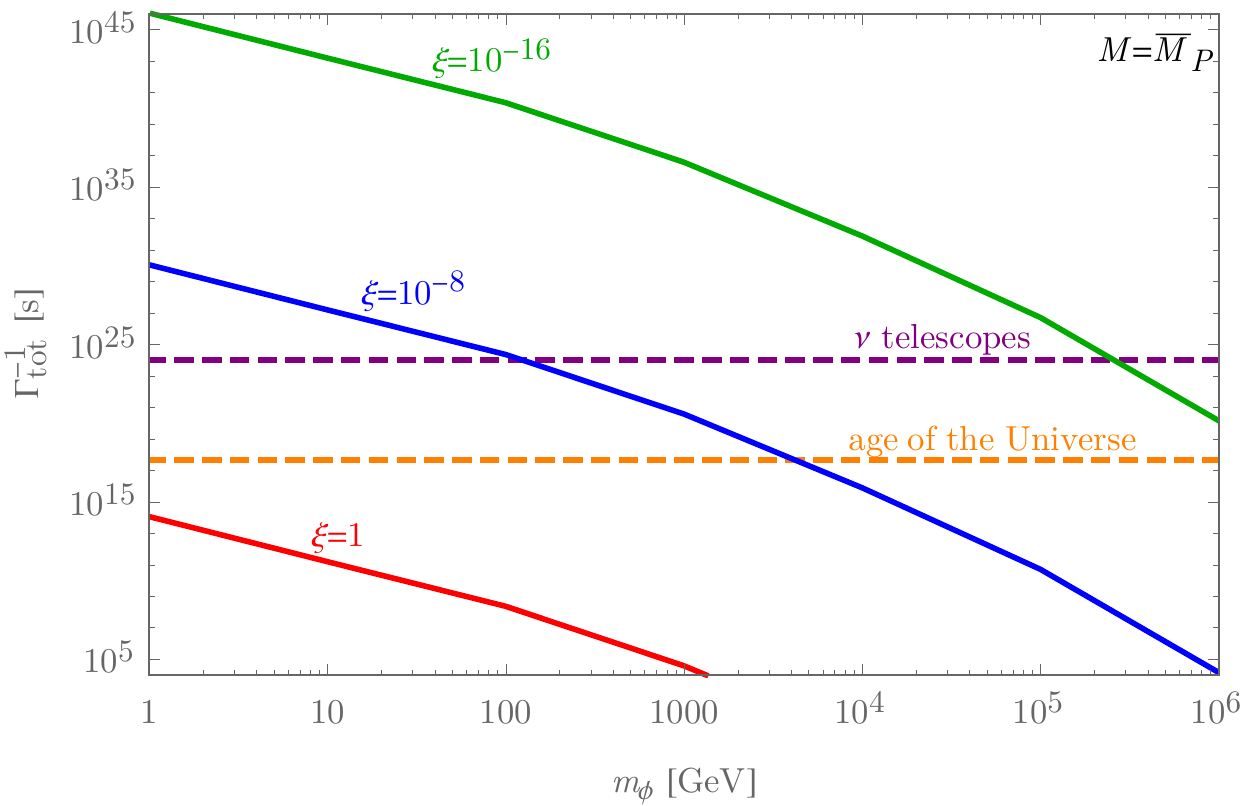} 
\end{center}
\caption{\small  {\it Left panel:} decay branching ratios for the scalar singlet dark matter candidate, $\phi$, as a function of its  mass. {\it Right panel:} Total inverse decay width assuming $M=\overline{M}_P$ for three reference values of the nonminimal coupling parameter $\xi$, compared to the age of the Universe and to a conservative lower limit on the total inverse width from observations of the cosmic neutrino flux.}
\label{fig:Sing} 
\end{figure}

The decay branching ratios of the scalar singlet dark matter candidate are depicted in the left panel of Fig.~\ref{fig:Sing}, where we have included only those decay modes with branching ratios above $5\%$.\footnote{All numerical results in this paper have been cross-checked against {\textsc{MadGraph$5_{-}$aMC@NLO}}~\cite{Alwall:2014hca}.} For low dark matter masses, between a few GeV and the electroweak scale, the decay rate is dominated by $\phi\to {\overline{q}}qg$ (a sum over flavors is implicitly understood). The decay $\phi\to{\overline{f}}f\gamma$ bears a relative $\alpha_{em}/\alpha_s$ suppression and can be safely neglected, while $\phi\to {\overline{f}}f$ is favored by phase space but is helicity suppressed, hence it has a significant effect only close to thresholds. For very large dark matter masses, on the other hand, the rate is dominated by the four-body decay $\phi\to (WW,ZZ)hh$. This process has a large phase space suppression which is nonetheless overturned for large masses by the $m_{\phi}^7/v^4$ scaling of the decay rate; we find numerically that this process dominates when $m_{\phi}\gtrsim {\cal{O}}(10^5)$ GeV. For intermediate masses, $v\lesssim m_{\phi}\lesssim 10^5$ GeV, the rate is dominated by the process $\phi\to Z(W){\overline{f}}f^{(\prime)}$, except in the rather small window $v\lesssim m_{\phi}\lesssim$ few TeV, where 2-body decays with high thresholds, namely $\phi\to WW,ZZ,hh,{\overline{t}}t$, are of comparable size to the phase-space suppressed three-body channels. Compared to our previous work~\cite{Cata:2016dsg}, in Fig.~\ref{fig:Sing} we have taken into account the running of $\alpha_s$; differences in the branching ratios are only substantial for $m_{\phi}$ of the order of a few GeV.     

Due to the different dependence of the various decay branching ratios on the dark matter mass, the total decay rate will usually be dominated by a few channels. This allows to derive an approximate lower bound on the total decay rate for a given value of the dark matter mass:
\begin{align}
\displaystyle{
\Gamma_{\phi}\gtrsim \frac{\xi^2M^2 m_\phi^3}{8\pi \overline M_P^{4}} \times \begin{cases}
 \displaystyle 2n_q\,\frac{\alpha_s}{\pi}, & m_\phi\sim 1-200  \GeV,   \vspace{0.2cm}\\
\displaystyle{1+2n_q\,\frac{\alpha_s}{\pi}},& m_\phi\sim 0.2-1 \TeV,  \vspace{0.2cm}\\
\displaystyle\frac{3}{(2\pi)^2}\frac{m_\phi^2}{v^2},& m_\phi\sim 1-100  \TeV, \vspace{0.2cm}\\
\displaystyle\frac{1}{10(8\pi)^4}\frac{m_\phi^4}{v^4}, & m_\phi\gtrsim 100  \TeV,
 \end{cases}}
\label{eq:total-Gamma-limitS}
\end{align}
where $n_q$ is the number of quarks kinematically accessible in the decay.

In Fig.~\ref{fig:Sing}, right panel, we show the predicted total inverse width as a function of the dark matter mass for three reference values of the nonminimal coupling parameter, $\xi=1,~10^{-8},~10^{-16}$, confronted with the age of the Universe and the conservative mass-independent lower limit $\Gamma^{-1}_{\rm tot}\gtrsim 10^{24}\, {\rm s}$ from measurements of the cosmic neutrino flux. The latter requirement in turn translates into the following upper limits on the nonminimal coupling parameter:
\begin{align}
\left|\xi\right| \lesssim  \begin{cases}
 \displaystyle 2\times10^{-8}\,\left(\frac{M}{\overline{M}_P}\right)^{-1}\left(\frac{m_\phi}{100 \GeV}\right)^{-3/2},  & m_\phi\sim 1-200  \GeV,  \vspace{0.3cm}\\
\displaystyle 8\times10^{-10}\,\left(\frac{M}{\overline{M}_P}\right)^{-1}\left(\frac{m_\phi}{500 \GeV}\right)^{-3/2},& m_\phi\sim 0.2-1  \TeV, \vspace{0.3cm} \\
\displaystyle 2\times10^{-14}\,\left(\frac{M}{\overline{M}_P}\right)^{-1}\left(\frac{m_\phi}{50 \TeV}\right)^{-5/2},&m_\phi\sim 1-100  \TeV, \vspace{0.3cm}\\
\displaystyle 4\times10^{-15}\,\left(\frac{M}{\overline{M}_P}\right)^{-1}\left(\frac{m_\phi}{100 \TeV}\right)^{-7/2},& m_\phi\gtrsim 100  \TeV.
 \end{cases}
\label{eq:xi-limitS}
\end{align}  
The strong suppression necessary for $\xi$ and/or $M$ calls for an explanation. The simplest possibility consists in postulating that the dark matter is charged under a gauge symmetry, such that the nonminimal coupling term $R \phi$ is forbidden altogether. This term, however, can be effectively generated upon symmetry breaking, as is the case for the inert doublet model which we discuss in the next section. 

\subsection{Inert doublet dark matter}
\label{sec:inert}
The inert doublet model~\cite{Deshpande:1977rw,LopezHonorez:2006gr} consists in extending the Standard Model particle content by one extra scalar doublet $\Phi$ with the same gauge quantum numbers as the Standard Model Higgs doublet. Furthermore, the model postulates a discrete $Z_2$ symmetry under which $\Phi$ is odd while all the Standard Model particles are even. With this assignment of gauge and discrete charges, the field $\Phi$ does not have tree-level interactions with any of the Standard Model fermions, but allows interactions with the Standard Model via the gauge and the Higgs portals. The Lagrangian of the model thus reduces to 
\begin{equation}
\mathcal L_{\DM}= g_{\mu\nu}(D^\mu \Phi)^\dagger (D^\nu \Phi) - V(\Phi,H)\;,
\end{equation}
where $V(\Phi,H)$ is the most general renormalizable $Z_2$-invariant potential involving the scalar fields $\Phi$ and $H$:
\begin{equation}
V(\Phi,H) = m_2^2\left|\Phi\right|^2  
+ \lambda_2\left|\Phi\right|^4
+\lambda_3\left|H\right|^2\left|\Phi\right|^2
+\lambda_4\left|H^\dagger \Phi\right|^2
+\dfrac{1}{2} \left( \lambda_5(H^\dagger \Phi)(H^\dagger \Phi)+{\rm \text{\small h.c.}}\right)\;.
\end{equation}
After electroweak symmetry breaking, the scalar fields can be cast as:
\begin{equation} \label{eq:scalar-fields}
 H =\frac{1}{\sqrt{2}}\begin{pmatrix} 0 \\  v+h\end{pmatrix} \;, \hspace{40pt} \Phi= \begin{pmatrix} \eta^+ \\ \frac{1}{\sqrt2} \left( \eta + i A \right) \end{pmatrix}  \;,
\end{equation}
where we implicitly chose to work in the unitary gauge. Here, $v=246$ GeV is the vacuum expectation value of the Higgs field. The multiplet $H$ contains only one physical scalar state, the Standard Model Higgs boson $h$, with $m_h=125$ GeV  \cite{Aad:2012tfa,Chatrchyan:2012xdj}. On the other hand, the $Z_2$-odd scalar sector contains one CP-even ($\eta$), one CP-odd ($A$) and two charged ($\eta^\pm$) scalar fields. 

In flat spacetime, the lightest particle of the $Z_2$-odd sector is absolutely stable; viable dark matter candidates of the model are then the CP-even scalar, $\eta$, and the CP-odd scalar, $A$. On the other hand, in the presence of gravitational interactions, $Z_2$-breaking may occur through operators where the doublet $\Phi$ couples nonminimally to gravity, potentially  leading to dark matter decay. The action in curved spacetime of the model, including the lowest dimensional nonminimal coupling operator, reads:
\begin{equation} \label{eq:CompleteActionJordaneta}
  \mathcal S = \int d^4x \sqrt{-g}\left[ -\frac{R }{2 \kappa^2}+ \mathcal L_{\rm SM}(X) + \mathcal L_{\DM}(\Phi,H) - \xi R (H^{\dagger}\Phi+\Phi^\dagger H)\right]\;.
\end{equation}
The Weyl factor in this case is 
\begin{align}
\Omega^2(\Phi,H)=1+2\kappa^2\xi(H^{\dagger}\Phi+\Phi^\dagger H)\;,
\end{align}
and the transition to the Einstein frame leads to 
\begin{align} \label{eq:scalardoubletActionEinstein}
  \mathcal S=\int d^4x \sqrt{-\widehat g}&\left( - \frac{\widehat R}{2 \kappa^2} + \frac{1}{2\Omega^2}\left(1 + \frac{6 (\partial_{\eta}\Omega)^2}{ \kappa^2} \right) {\widehat g}_{\mu\nu}( D^\mu \eta)(D^\nu \eta)  
\right. \nonumber \\
  &\left.
+\frac{6(\partial_{\eta}\Omega)(\partial_h\Omega)}{\kappa^2\Omega^2}\widehat{g}_{\mu\nu}D^{\mu}\eta D^{\nu}h 
+\frac{1}{2\Omega^2} {\widehat g}_{\mu\nu}( D^\mu A)(D^\nu A)
\right. \nonumber \\
  & \left. 
+\frac{1}{\Omega^2} {\widehat g}_{\mu\nu}( D^\mu \eta^+)(D^\nu \eta^-)
+\frac{3 (\partial_h \Omega)^2}{ \kappa^2 \Omega^2} {\widehat g}_{\mu\nu}( D^\mu h)(D^\nu h)
\right. \nonumber \\
  & \left.
- \frac{1}{\Omega^4} V(\Phi,H) + \mathcal {\widehat{L}}_\text{\rm SM} \right).
\end{align}

We are mostly interested in $\mathcal {\widehat{L}}_\text{\rm SM}$, which contains the couplings of the Standard Model to dark matter through different powers of the Weyl factor. In the unitary gauge it reduces to
\begin{align}
\Omega^2(\eta,h)=1+2\kappa^2\xi(v+h)\eta\;.
\end{align}
We note that $\Omega$ does not depend on $A$. Therefore, the CP-odd dark matter candidate is absolutely stable even in the presence of a $Z_2$ breaking term proportional to the Ricci scalar. In the following we will examine the decay through the gravity portal of the CP-even scalar dark matter candidate, $\eta$. This scenario has close similarities with the scalar singlet case already examined in the previous section, hence we will just highlight the aspects where they differ.
 
The interactions inducing dark matter decay can be determined by expanding ${\mathcal{\widehat{L}}}_{\rm SM}$ to linear order in $\eta$. The result reads
\begin{align} \label{eq:LSM_Doublet_DM_Einstein}
{\mathcal{\widehat{L}}}_{{\rm SM},\eta}&=-2\kappa^2\xi(v+h)\eta\left[\frac{3}{2}{\mathcal{\widehat T}}_{f}+{\mathcal{\widehat T}}_{H}+2({\mathcal{L}}_{Y}-V_{H}) \right]\;.
\end{align}
Notice that, as compared to the scalar singlet case, the mass parameter $M$ is here identified with the Higgs vacuum expectation value. The small factor $v/{\overline{M}}_P\ll 1$ will then provide a natural suppression of the decay rate, even for $\xi\sim{\cal O}(1)$.

Another important difference with respect to the singlet case is the increase in the number of decay modes. This is prompted by the presence of the Higgs boson already in the Weyl factor. As a result, up to five-body decays will have to be examined. The full list of vertices can be found in Appendix \ref{app:rules}. In Table~\ref{tab:etaDecays} we list the kinematical dependence of the partial decay dates, omitting the common prefactor $\xi^2 v^2\kappa^4$ and the phase space factors.
\begin{table}[t!]
\centering
\renewcommand{\arraystretch}{1.2}
  \begin{tabular}{| l | c | |l | c |}
    \hline
    \hskip 1.0cm Decay mode &  Scaling & \hskip 1.0cm Decay mode &  Scaling \\
    \hline
    $\eta\rightarrow hh, WW, ZZ$                          & $m_\eta^3$  & & \\
    $\eta\rightarrow f\overline f$		                        & $m_f^2 m_\eta$ & & \\
    \hline
    $\eta\rightarrow hhh$			                        & $m_\eta^5/v^2$ & $\eta\rightarrow hhhhh$			                        & $m_\eta^5/v^2$\\
    $\eta\rightarrow WWh, ZZh$	                        & $m_\eta^5/v^2$ & $\eta\rightarrow WWhhh, ZZhhh$			                        & $m_\eta^9/v^6$ \\
    $\eta\rightarrow f\overline fh$		                        & $m_f^2 m_\eta^3/v^2$ & $\eta\rightarrow f\overline fhh$				                        & $m_f^2m_\eta^5/v^4$ \\
    $\eta\rightarrow f\overline f'W, f\overline fZ$	& $m_\eta^5/v^2$ & $\eta\rightarrow f\overline f'Wh, f\overline fZh$		                        & $m_\eta^7/v^4$ \\
    $\eta\rightarrow f\overline f\gamma, q\overline qg$	& $m_\eta^3$ & $\eta\rightarrow f\overline f\gamma h, q\overline qg h$			                        & $m_\eta^5/v^2$ \\
    \hline
    $\eta\rightarrow hhhh$			                        & $m_\eta^3$ & & \\
    $\eta\rightarrow WWhh, ZZhh$			& $m_\eta^7/v^4$ & & \\
    \hline
  \end{tabular}
  \caption{Same as Table \ref{tab:phiDecays}, but for the inert doublet dark matter candidate $\eta$. 
In the left-hand-side panel we list the decay modes common to the scalar singlet and in the right-hand-side panel the new decay modes.}
  \label{tab:etaDecays}
\end{table}

Exact analytical expressions can be straightforwardly obtained for the two-body decays:
\begin{align}
\Gamma_{\eta\to hh}&=\frac{\xi^2v^2\kappa^4}{32\pi}m_{\eta}^3(1+2x_h)^2(1-4x_h)^{1/2}\;,\nonumber \\
\Gamma_{\eta\to ZZ}&=\frac{\xi^2v^2\kappa^4}{32\pi}m_{\eta}^3(1-4x_Z+12x_Z^2)(1-4x_Z)^{1/2}\;,\nonumber\\
\Gamma_{\eta\to WW}&=\frac{\xi^2v^2\kappa^4}{16\pi}m_{\eta}^3(1-4x_W+12x_W^2)(1-4x_W)^{1/2}\;,\nonumber \\
\Gamma_{\eta \to \overline{f}_{\!i} f_i}&=N_c^{(f_i)}\frac{\xi^2v^2\kappa^4}{8\pi}m_{\eta}^3\,x_{\! f_i}(1-4x_{\!f_{i}})^{3/2}\;.
\end{align} 

\begin{figure}[t!]
\begin{center}
 \includegraphics[width=0.49\textwidth]{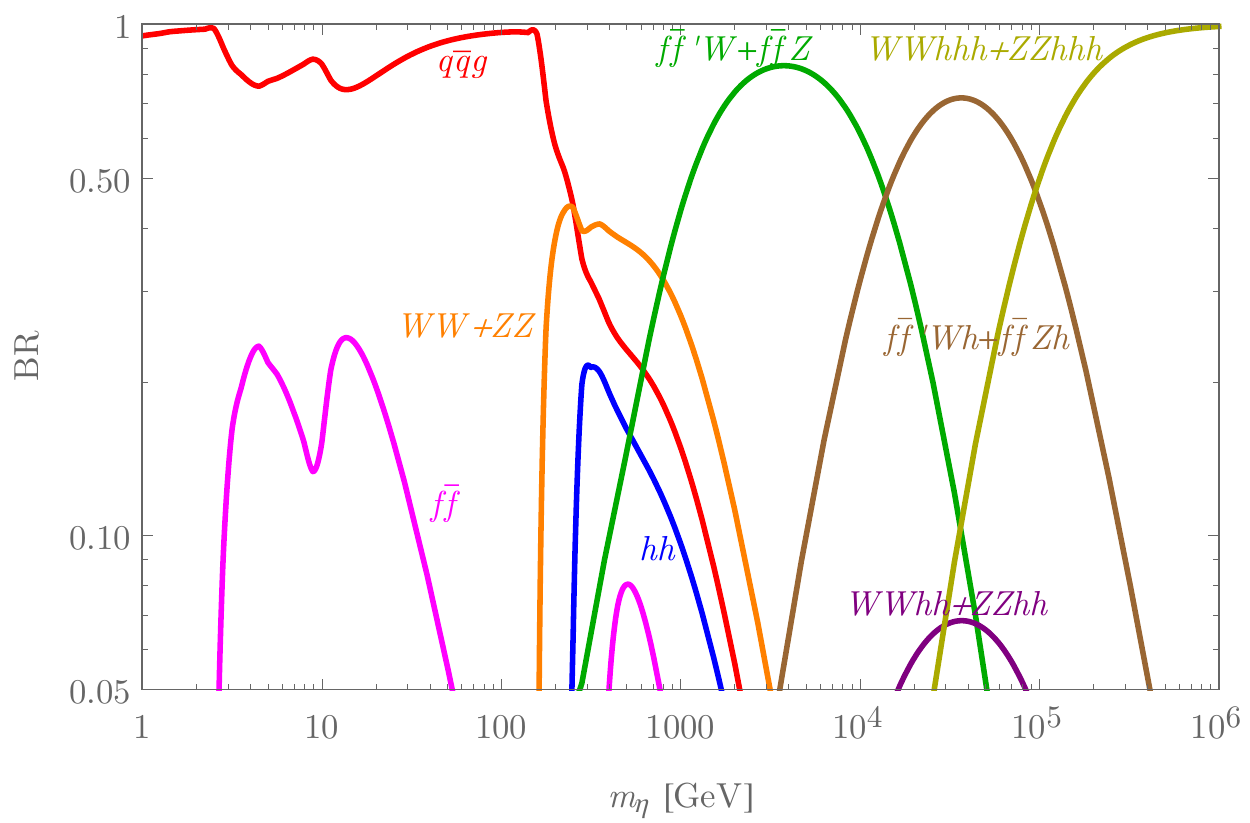} 
 \includegraphics[width=0.49\textwidth]{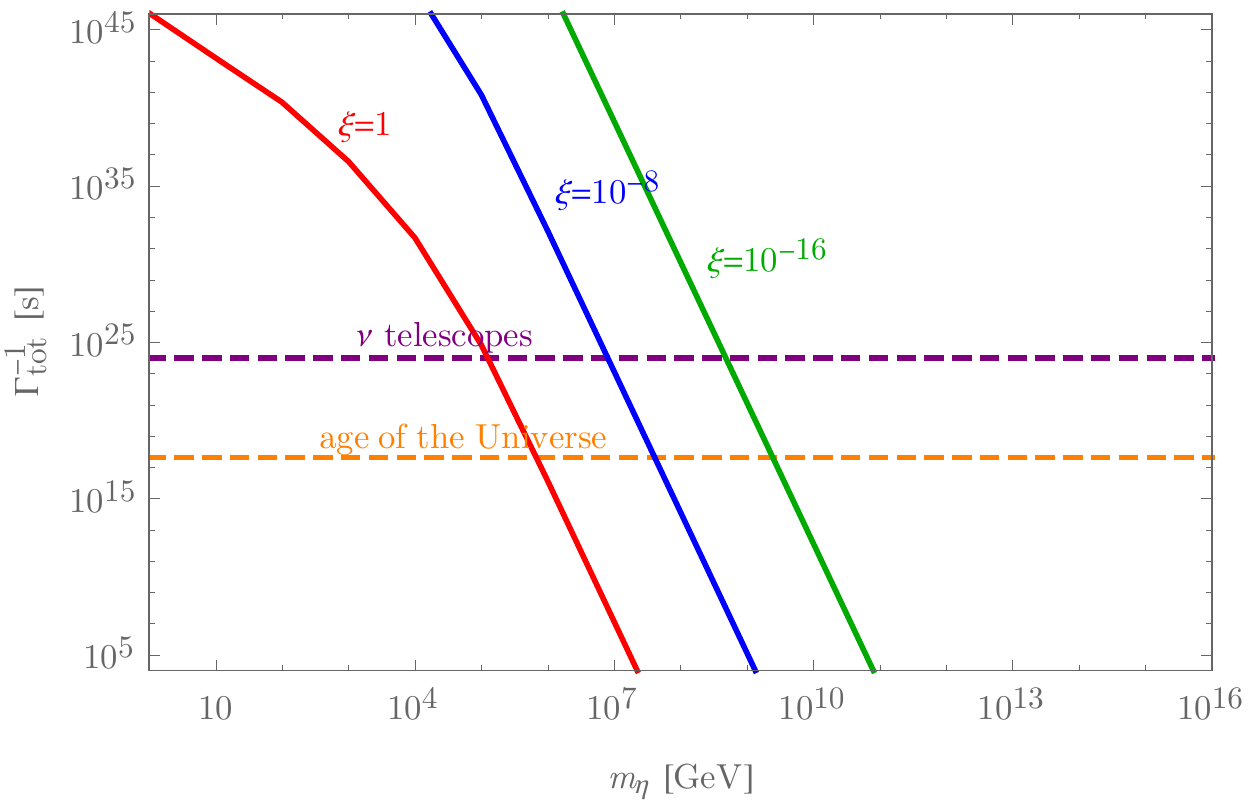} 
\end{center}
\caption{\small  Same as Fig.~\ref{fig:Sing}, but for the inert doublet dark matter candidate, $\eta$. }
\label{fig:Doublet} 
\end{figure}

For the three-, four- and five-body decay rates, the analytical expressions are rather complicated, although they greatly simplify in the limit $m_{\eta}\gg m_{\rm SM}$: 
\begin{align}
\Gamma_{\eta\to \overline{q}_i q_i g}&\simeq \alpha_s\frac{\xi^2v^2\kappa^4}{4\pi^2}m_{\eta}^3\;,\nonumber\\
\Gamma_{\eta\to {\overline{f}_{\!i}}f_j^\prime W}&\simeq \frac{3}{4\sqrt{2}}G_FN_c^{(f_i)}|U_{ij}|^2 \frac{\xi^2v^2\kappa^4}{(4\pi)^3}m_{\eta}^5\;,\nonumber\\
\Gamma_{\eta\to {\overline{f}}_{\!i} f_i Z}&\simeq \frac{3}{2\sqrt{2}}G_FN_c^{(f_i)}(g_V^2+g_A^2)\frac{\xi^2v^2\kappa^4}{(4\pi)^3}m_{\eta}^5\;,\nonumber\\
\Gamma_{\eta\to {\overline{f}_{\! i}}f_j^\prime Wh}&\simeq \frac{3\sqrt{2}}{160}G_FN_c^{(f_i)}|U_{ij}|^2 \frac{\xi^2\kappa^4}{(4\pi)^5}m_{\eta}^7\;,\nonumber\\
\Gamma_{\eta\to {\overline{f}_{\! i}}f_i Zh}&\simeq \frac{3\sqrt{2}}{80}G_FN_c^{(f_i)}(g_V^2+g_A^2)\frac{\xi^2\kappa^4}{(4\pi)^5}m_{\eta}^7\;,\nonumber\\
\Gamma_{\eta\to WWhhh}&\simeq \frac{2\xi^2\kappa^4}{75(8\pi)^7v^4}m_{\eta}^9\;,\nonumber\\
\Gamma_{\eta\to ZZhhh}&\simeq \frac{\xi^2\kappa^4}{ 75(8\pi)^7v^4}m_{\eta}^9\;.
\end{align}
The exact decay branching ratios are plotted in Fig.~\ref{fig:Doublet}.  For dark matter masses below $4$ TeV, the branching ratios are qualitatively the same as in the singlet case. However, for masses above that value, the existence of higher final-state multiplicities changes the picture: $WWhhh$ final states are asymptotically dominant starting at roughly $100$ TeV, while $WWhh$ are practically negligible. From $m_{\eta}\sim (1-100)$ TeV, $Z(W){\overline{f}}f^{(\prime)}$ dominate until taken over by $Z(W){\overline{f}}f^{(\prime)}h$ decays.

The total decay width can be approximated by:
\begin{align}
\displaystyle{
\Gamma_\eta\gtrsim \frac{\xi^2 v^2 m_\eta^3}{8\pi \overline{M}_P^4} \times \begin{cases}
 \displaystyle 2n_q\,\frac{\alpha_s}{\pi}, & m_\eta\sim 1-200  \GeV,   \vspace{0.2cm}\\
\displaystyle{1+2n_q\,\frac{\alpha_s}{\pi}},& m_\eta\sim 0.2-1 \TeV,  \vspace{0.2cm}\\
\displaystyle\frac{3}{(2\pi)^2}\frac{m_\eta^2}{v^2},& m_\eta\sim 1-10  \TeV, \vspace{0.2cm}\\
\displaystyle \frac{3}{5(4\pi)^4}\frac{m_\eta^4}{v^4}, & m_\eta\sim 10-100  \TeV,   \vspace{0.2cm}\\
\displaystyle\frac{1}{ 25(8\pi)^6}\frac{m_\eta^6}{v^6}, & m_\eta\gtrsim 100  \TeV,
 \end{cases}}
\label{eq:total-Gamma-limitD}
\end{align}
and, due to the small factor $v/{\overline{M}}_P\ll 1$, is expected to be significantly smaller than for the real scalar dark matter scenario discussed in Section \ref{sec:scalar}.

The requirement $\Gamma^{-1}_{\rm tot} \gtrsim 10^{24}\,\text{s}$ then translates into the following bounds for the parameter $\xi$:
\begin{align}
|\xi|  \lesssim  \begin{cases}
 \displaystyle 2\times10^{8}\,\left(\frac{m_\eta}{100 \GeV}\right)^{-3/2},  & m_\eta\sim 1-200  \GeV,  \vspace{0.3cm}\\
\displaystyle 8\times10^{6}\,\left(\frac{m_\eta}{500 \GeV}\right)^{-3/2},& m_\eta\sim 0.2-1  \TeV, \vspace{0.3cm} \\
\displaystyle 5\times10^{4}\,\left(\frac{m_\eta}{5 \TeV}\right)^{-5/2},&m_\eta\sim 1-10  \TeV, \vspace{0.3cm}\\
\displaystyle 40\,\left(\frac{m_\eta}{50 \TeV}\right)^{-7/2},&m_\eta\sim 10-100  \TeV, \vspace{0.3cm}\\
\displaystyle 3\,\left(\frac{m_\eta}{100 \TeV}\right)^{-9/2},& m_\eta\gtrsim 100  \TeV.
 \end{cases}
\label{eq:xi-limitD}
\end{align}

In Fig.~\ref{fig:Doublet}, right panel, we plot the exact dependence of the dark matter lifetime on its mass for different reference values of $\xi$. We find that for $\xi={\cal O}(1)$ the gravity portal induces too fast dark matter decay when $m_\eta\gtrsim 100$ TeV. Given the previous results, it is tantalizing to speculate that the high energy neutrino events detected by IceCube~\cite{Aartsen:2013bka,Aartsen:2014gkd} might have their origin in the decay via the gravity portal of inert doublet scalars, with masses of a few PeV and $\xi\sim 10^{-7}-10^{-3}$. This scenario, on the other hand, would require a non-thermal production mechanism to comply with the unitarity limits on the mass of thermal relics \cite{Griest:1989wd}. A detailed spectral and morphological analysis of the signal is out of the scope of this paper (see~\cite{Feldstein:2013kka,Esmaili:2013gha} for some recent analyses).  

\subsection{Fermionic singlet dark matter}
\label{sec:fermionic}
Finally, we consider a fermionic dark matter candidate, $\chi$, singlet under the Standard Model gauge group and transforming non-trivially under a global symmetry~\cite{Kim:2006af}. The dark matter Lagrangian simply reads:
\begin{align}
\mathcal{L}_{\DM}=\frac{i}{2}\,{\overline{\chi}}\!\!\stackrel{\longleftrightarrow}{\slash{\!\!\!\!\nabla}}\!\!\!\chi-m_{\chi}\overline{\chi}\chi+V(\chi,X)\;,
\end{align}
where $\slash{\!\!\!\!\nabla}=\gamma^a e^{\mu}_a\nabla_{\mu}$, with $\nabla_\mu=\partial_\mu -\frac{i}{4}e_\nu^b (\partial_\mu e^{\nu c}) \sigma_{bc}$, and $V(\chi,X)$ is an interaction term with the Standard Model. Fermionic singlet dark matter candidates have been extensively discussed in the literature in conjunction with a scalar mediator. These frameworks assume that both the dark matter and the mediator are odd under a discrete $Z_2$ symmetry, while all Standard Model fields are even, and that the gauge charges allow a Yukawa interaction term with one Standard Model field, which acts as a ``portal''. For appropriate values of the model parameters, the fermionic singlet abundance from thermal freeze-out matches the observed dark matter abundance. Furthermore, the models offer a rich phenomenology in direct, indirect and collider searches (for a review, see~\cite{Garny:2015wea}).

The embedding of the model in curved spacetime in general contains a $Z_2$-breaking term proportional to the Ricci scalar and linear in the fermionic singlet field. It can be checked that the lowest dimensional operator of this class is dimension 6, i.e.
\begin{equation} \label{eq:Jordanfermion}
  \mathcal S = \int d^4x \sqrt{-g}\left[ -\frac{R }{2 \kappa^2}+ \mathcal L_{\rm SM}(X) + \mathcal L_{\DM}(\chi,H) - \frac{\xi_i}{M^2} R \left( \overline \ell_i \widetilde{H} \chi + \overline\chi\widetilde{H}^{\dagger}\ell_i \right)\right]\;,
\end{equation}
where $\widetilde{H}=i\sigma_2 H^*$, $\ell_i$ stands for an electroweak lepton doublet with flavor $i$, and $M$ is the energy scale of the new physics generating the dimension-6 operator.\footnote{We have considered a Dirac fermion dark matter; for a Majorana fermion dark matter the only difference is that all the rates should be multiplied by a factor of 2.} A simple ultraviolet completion of Eq.~(\ref{eq:Jordanfermion}) arises when the model also includes a $Z_2$-odd scalar, $S$, with mass $m_S$ larger than the mass of the fermionic singlet, along the lines of the models discussed above. Among other terms, the potential contains the interaction 
\begin{align}
V(\chi,\ell;S)\supset \lambda_{S,i}\,{\overline{\ell}_i}\tilde{S}\chi + {\mathrm{h.c.}}
\end{align}
In flat spacetime $S$ can decay into $\chi$ plus a Standard Model neutrino, but $\chi$ is absolutely stable. The embedding into curved spacetime may contain a nonminimal coupling term between the $Z_2$-odd scalar $S$ and the curvature, of the form $\xi_{SH} R(H^{\dagger} S+ S^{\dagger} H)$. At energy scales much below $m_S$, the scalar $S$ can be integrated out and the model is well described by an effective Lagrangian of the form Eq.~(\ref{eq:Jordanfermion}), with 
\begin{align}
\frac{\xi_i}{M^2}=-\frac{\lambda_{S,i}\,\xi_{SH}}{m_S^2}\;.
\end{align}
The higher dimensionality of the  dominant nonminimal operator to gravity implies an additional suppression of the decay rate by a power of $M^4$ in addition to $\overline M_P^4$, which in turn translates into smaller decay rates compared to the scalar dark matter candidates when $M\gg m_\chi$.

\begin{table}[t!]
\centering
\renewcommand{\arraystretch}{1.2}
  \begin{tabular}{| l | c | |l | c |}
    \hline
    \hskip 1.0cm Decay mode &  Scaling & \hskip 1.0cm Decay mode &  Scaling \\
    \hline
    $\chi\rightarrow hh\nu, WW\nu, ZZ\nu$                          & $m_\chi^7$  & & \\
    $\chi\rightarrow f\overline f\nu$		                        & $m_f^2 m_\chi^5$ & & \\
    \hline
    $\chi\rightarrow hhh\nu$			                        & $m_\chi^9/v^2$ & $\chi\rightarrow hhhhh\nu$			                        & $m_\chi^9/v^2$\\
    $\chi\rightarrow WWh\nu, ZZh\nu$	                        & $m_\chi^9/v^2$ & $\chi\rightarrow WWhhh\nu, ZZhhh\nu$			                        & $m_\chi^{13}/v^6$ \\
    $\chi\rightarrow f\overline fh\nu$		                        & $m_f^2 m_\chi^7/v^2$ & $\chi\rightarrow f\overline fhh\nu$				                        & $m_f^2m_\chi^9/v^4$ \\
    $\chi\rightarrow f\overline f'W\nu, f\overline fZ\nu$	& $m_\chi^9/v^2$ & $\chi\rightarrow f\overline f'Wh\nu, f\overline fZh\nu$		                        & $m_\chi^{11}/v^4$ \\
    $\chi\rightarrow f\overline f\gamma\nu, q\overline qg\nu$	& $m_\chi^7$ & $\chi\rightarrow f\overline f\gamma h\nu, q\overline qg h\nu$			                        & $m_\chi^9/v^2$ \\
    \hline
    $\chi\rightarrow hhhh\nu$			                        & $m_\chi^7$ & & \\
    $\chi\rightarrow WWhh\nu, ZZhh\nu$			& $m_\chi^{11}/v^4$ & & \\
    \hline
  \end{tabular}
  \caption{Same as Table \ref{tab:phiDecays}, but for the fermionic singlet dark matter candidate $\chi$.}
  \label{tab:chiDecays}
\end{table}

To determine the decay vertices we Weyl-transform the Lagrangian, focusing on the Standard Model part. This is given by Eq.~(\ref{eq:SMLagrangianEinstein}), identifying the Weyl factor as
\begin{align}
\Omega^2(\chi,\nu,h)=1+\sqrt{2}\kappa^2M^{-2}\xi_i(v+h)(\overline{\nu}_{L,i}\chi+\overline\chi \nu_{L,i})\;.
\end{align}
Expanding to linear order in the dark matter field one  finally obtains  the relevant interaction Lagrangian:
\begin{align}\label{eq:fermionFR}
{\mathcal{\widehat L}}_{{\rm SM},\chi}&= -\sqrt{2}\frac{\kappa^2}{M^2}\xi_i (v+h)(\overline\nu_{L,i}\chi+\overline{\chi}\nu_{L,i})\left[\frac{3}{2}{\mathcal{\widehat T}}_{f}+{\mathcal{\widehat T}}_{H}+2({\mathcal{L}}_{Y}-V_{H}) \right]\;.
\end{align}

From the structure of the interaction, one finds that the full list of vertices can be derived from those in the inert doublet dark matter scenario, appending to each decay amplitude the fermionic current $\frac{1}{2\sqrt{2}M^2}{\overline{u}}(p_{\nu})(1-\gamma_5)u(p_{\chi})$. Hence, up to six-body decays ought to be considered, especially for large values of the dark matter mass, where these modes are expected to dominate.

\begin{figure}[t]
\begin{center}
 \includegraphics[width=0.49\textwidth]{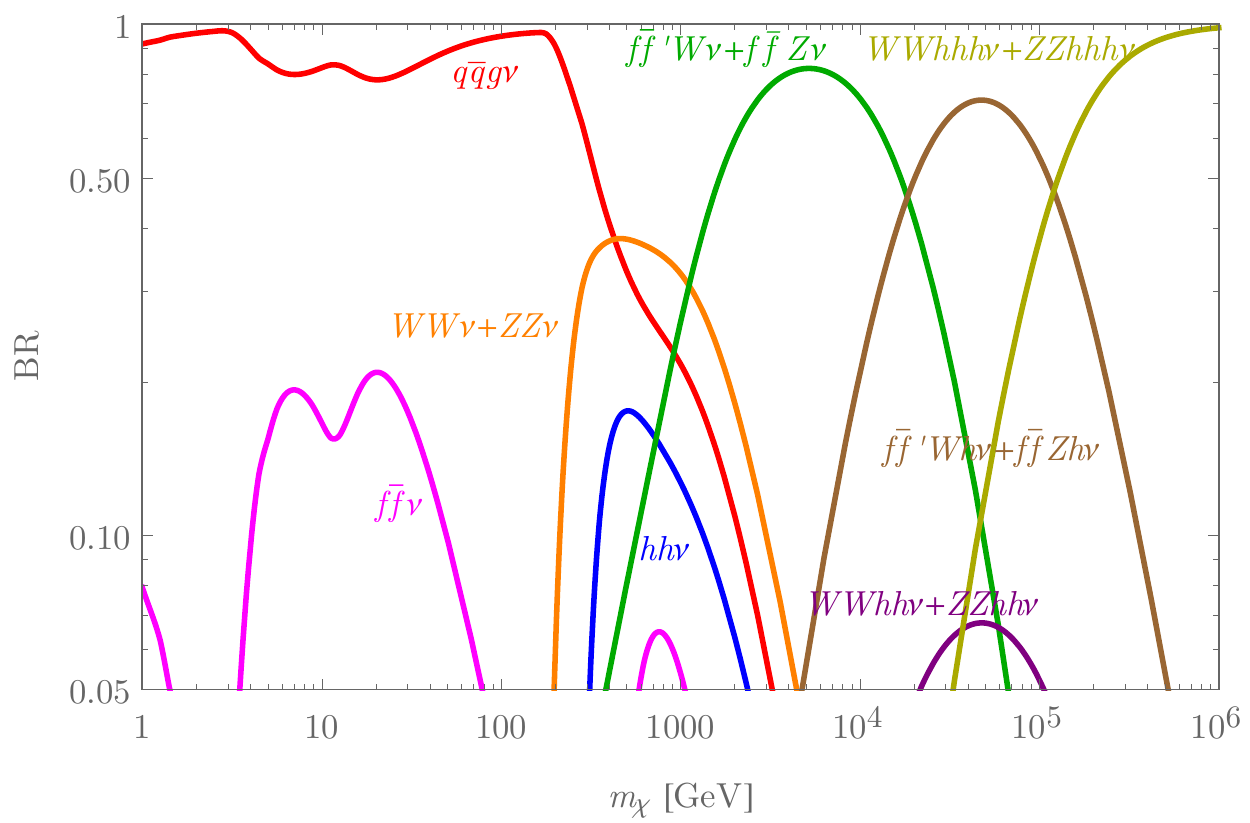} 
 \includegraphics[width=0.49\textwidth]{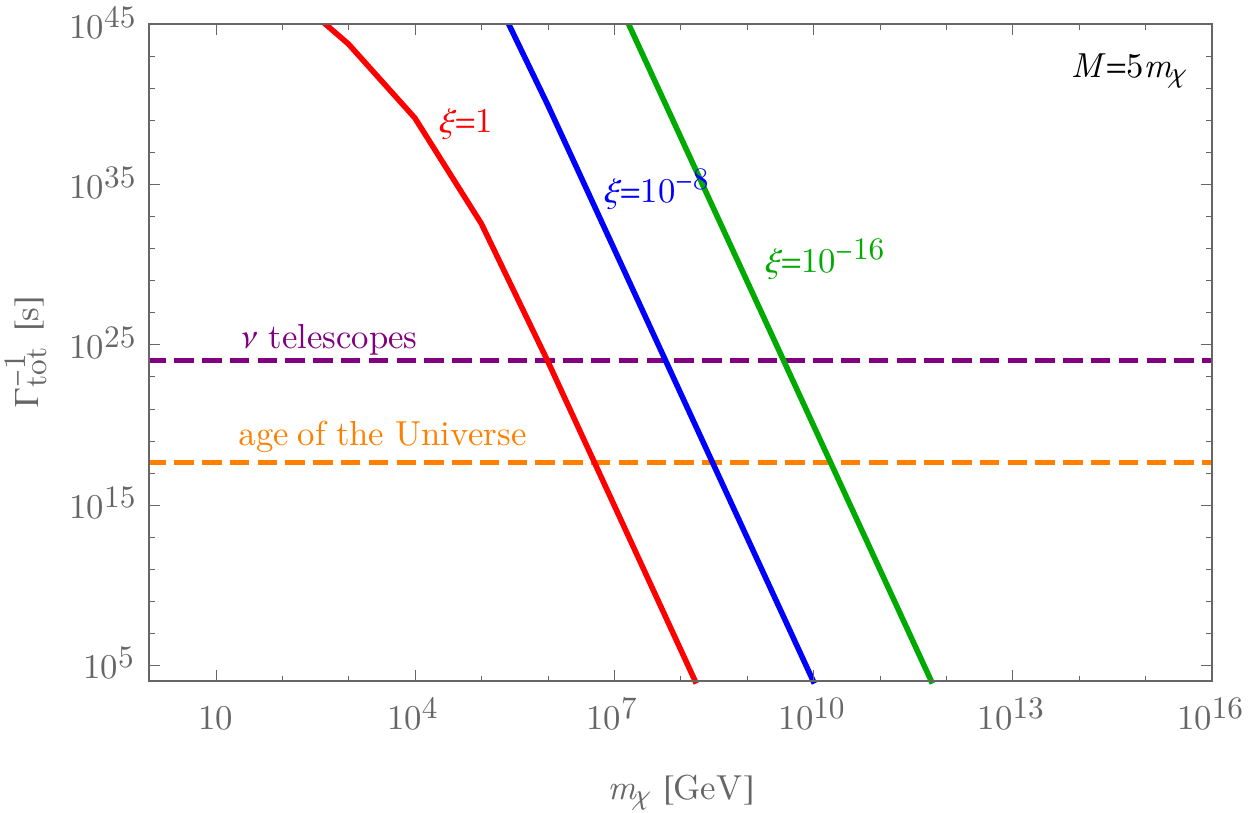} 
\end{center}
\caption {\small Same as Fig.~\ref{fig:Sing}, but for the fermionic singlet dark matter candidate, $\chi$, assuming $M=5 m_\chi$.}
\label{fig:FermSing} 
\end{figure}

In Table~\ref{tab:chiDecays} we list the dependence of the different decay modes on dark matter and Standard Model masses, omitting  the common prefactor $\xi^2 M^{-4}\kappa^4 v^2$ and the phase space factors. Despite the different scaling of the separate decay modes when compared with the scalar doublet case, the branching ratios present a qualitatively similar picture, as seen in Fig.~\ref{fig:FermSing}.

Exact analytical expressions can be obtained for the three-body decay rates:
\begin{align}
\Gamma_{\chi\to hh\nu}&=\frac{\xi^2v^2\kappa^4}{15M^4(16\pi)^3}m_{\chi}^7g_1(x_h)\;, \nonumber \\
\Gamma_{\chi\to ZZ\nu}&=\frac{\xi^2v^2\kappa^4}{15M^4(16\pi)^3}m_{\chi}^7g_2(x_Z)\;, \nonumber\\
\Gamma_{\chi\to WW\nu}&=\frac{2\xi^2v^2\kappa^4}{15M^4(16\pi)^3}m_{\chi}^7g_2(x_W)\;, \nonumber\\
\Gamma_{\chi\to  \overline{f}_{\!i} f_i \nu}&=2N_c^{(f_i)}\frac{\xi^2v^2\kappa^4}{3M^4(16\pi)^3}\, x_{\! f_i}m_{\chi}^7g_3(x_{\! f_i})\;.
\end{align}
Here, we have summed the rates for all three flavors of the final state neutrino and we have defined $\xi^2 \equiv \sum_i \xi^2_i$. Besides,
\begin{align}
g_1(x)&=(1+7x-44x^2+810x^3-1260x^4)(1-4x)^{1/2}-120x^3(5-17x+21x^2)\log f(x)\;, \nonumber\\
g_2(x)&=(1-13x+156x^2+450x^3-540x^4)(1-4x)^{1/2}-120x^3(5-9x+9x^2)\log f(x)\;,\nonumber\\
g_3(x)&=(1-22x-42x^2+36x^3)(1-4x)^{1/2}+24x^2(3-4x+3x^2)\log f(x)\;,
\end{align}
with
\begin{align}
f(x)=\frac{1-2x+(1-4x)^{1/2}}{2x}\;.
\end{align}

The analytical expressions for the four-, five- and six-body decay rates are complicated, and we only show here approximate expressions in the limit $m_{\chi}\gg m_{\rm SM}$:
\begin{align}
\Gamma_{\chi\to \overline{q}_i q_i g  \nu}&\simeq \frac{16}{25}\alpha_s\frac{\xi^2v^2\kappa^4}{M^4(8\pi)^4}m_{\chi}^7\;,\nonumber \\
\Gamma_{\chi\to \overline{f}_{\!i} f_j^\prime W  \nu}&\simeq \frac{\sqrt{2}}{4}G_FN_c^{(f_i)}|U_{ij}|^2\frac{\xi^2v^2\kappa^4}{5(8\pi)^5M^4}m_{\chi}^9\;,\nonumber\\
\Gamma_{\chi\to \overline{f}_{\!i} f_i Z \nu}&\simeq G_FN_c^{(f_i)}(g_V^2+g_A^2)\frac{\xi^2v^2\kappa^4}{5\sqrt{2}(8\pi)^5M^4}m_{\chi}^9\;,\nonumber\\
\Gamma_{\chi\to \overline{f}_{\! i} f_j^\prime Wh \nu}&\simeq \sqrt{2}G_FN_c^{(f_i)}|U_{ij}|^2\frac{\xi^2\kappa^4}{175(8\pi)^7M^4}m_{\chi}^{11}\;,\nonumber\\
\Gamma_{\chi\to \overline{f}_{\! i} f_i Zh \nu}&\simeq 2\sqrt{2}G_FN_c^{(f_i)}(g_V^2+g_A^2)\frac{\xi^2\kappa^4}{175(8\pi)^7M^4}m_{\chi}^{11}\;,\nonumber\\
\Gamma_{\chi\to WWhhh \nu}&\simeq\frac{\xi^2\kappa^4}{6300(8\pi)^9v^4M^4}m_{\chi}^{13}\;,\nonumber \\
\Gamma_{\chi\to ZZhhh \nu}&\simeq \frac{\xi^2\kappa^4}{12600(8\pi)^9v^4M^4}m_{\chi}^{13}\;.
\end{align} 

The total decay width is then bounded from below by
\begin{align}
\displaystyle{
\Gamma_{\chi}\gtrsim \frac{4\xi^2v^2 m_\chi^7}{15(16\pi)^3M^4\overline M^4_P} \times \begin{cases}
 \displaystyle \frac{12}{5}n_q\,\frac{\alpha_s}{\pi}, & m_\chi\sim 1-200  \GeV,   \vspace{0.2cm}\\
\displaystyle{1+\frac{12}{5}n_q\,\frac{\alpha_s}{\pi}},& m_\chi\sim 0.2-1 \TeV,  \vspace{0.2cm}\\
\displaystyle \frac{3}{8\pi^2}\frac{m_{\chi}^2}{v^2}, & m_\chi\sim 1-10  \TeV,   \vspace{0.2cm}\\
\displaystyle\frac{6}{35(4\pi)^4}\frac{m_\chi^4}{v^4},& m_\chi\sim 10-100  \TeV, \vspace{0.2cm}\\
\displaystyle\frac{1}{140(8\pi)^6}\frac{m_\chi^6}{v^6}, & m_\chi\gtrsim 100  \TeV.
 \end{cases}}
\label{eq:total-Gamma-limitF}
\end{align}
Conservatively requiring $\Gamma^{-1}_{\rm tot}\gtrsim 10^{24}$~s, one finds the following upper limit on the nonminimal coupling parameter:
\begin{align}
\left|\xi \right| \lesssim  \begin{cases}
 \displaystyle 5\times10^{11}\,\left(\frac{M/m_\chi}{5}\right)^{2}\,\left(\frac{m_\chi}{100 \GeV}\right)^{-3/2},  & m_\chi\sim 1-200  \GeV,  \vspace{0.3cm}\\
\displaystyle 3\times 10^{10}\,\left(\frac{M/m_\chi}{5}\right)^{2}\,\left(\frac{m_\chi}{500 \GeV}\right)^{-3/2},& m_\chi\sim 0.2-1  \TeV, \vspace{0.3cm} \\
\displaystyle 2\times10^{8}\,\left(\frac{M/m_\chi}{5}\right)^{2}\,\left(\frac{m_\chi}{5 \TeV}\right)^{-5/2},&m_\chi\sim 1-10  \TeV, \vspace{0.3cm}\\
\displaystyle 3\times10^{5}\,\left(\frac{M/m_\chi}{5}\right)^{2}\,\left(\frac{m_\chi}{50 \TeV}\right)^{-7/2},&m_\chi\sim 10-100  \TeV, \vspace{0.3cm}\\
\displaystyle 3\times 10^{4}\,\left(\frac{M/m_\chi}{5}\right)^{2}\,\left(\frac{m_\chi}{100 \TeV}\right)^{-9/2},& m_\chi\gtrsim 100  \TeV.
 \end{cases}
\label{eq:xi-limitF}
\end{align}
These upper limits are less stringent than for the scalar dark matter decays due to the extra suppressing $m_{\chi}/M$ factors. 

In Fig.~\ref{fig:FermSing}, right panel, we plot the exact value for the total inverse width in terms of $m_{\chi}$ for different values of the parameter $\xi$, assuming $M=5 m_\chi$. As seen from the figure, compatibility with observations requires $m_\chi\lesssim 10^6$ GeV for $\xi={\cal O}(1)$, and $m_\chi\lesssim 10^{10}$ GeV for $\xi={\cal O}(10^{-16})$. Increasing the suppression scale $M$ of the dimension-six operator leads to a further suppression of the rate and heavier and heavier masses become allowed; for $M=\overline{M}_P$ the upper limit on the mass becomes $m_\chi\lesssim 10^{10}$ GeV.

We stress that these conclusions are only valid when the new physics generating the nonminimal coupling of the fermionic singlet dark matter to gravity is much heavier than the dark matter mass, such that the effective Lagrangian in Eq.~(\ref{eq:Jordanfermion}) provides a good description of the theory. In some other scenarios, in contrast, this condition might not hold. An interesting example is the class of models sketched at the beginning of this section, which also include a $Z_2$-odd scalar coupled nonminimally to gravity, as well as to the fermionic singlet and lepton doublet through a Yukawa term. The scenario where the scalar mediator is quasi-degenerate in mass with the dark matter presents important qualitative differences compared to the hierarchical scenario.  Most notably, the presence of the scalar mediator in the thermal plasma at the epoch of freeze-out can significantly affect the relic dark matter abundance due to coannihilations~\cite{Griest:1990kh}. Also, higher order processes, such as the annihilation into a fermion-antifermion pair with the associated emission of a gauge boson, can play an important role in indirect searches~\cite{Flores:1989ru,Bergstrom:1989jr,Bringmann:2007nk}.
More specifically, dark matter decays via the s-channel exchange of the scalar mediator, with a rate which is resonantly enhanced by the small difference between the dark matter mass and the mediator mass. Decays through the gravitational portal could then offer another window to indirectly detect dark matter particles produced via thermal freeze-out in the coannihilation regime. In this regime, the scalar mediator ought not to be integrated out and Eq.~(\ref{eq:Jordanfermion}) does not provide a good description of the model. A detailed analysis will be presented elsewhere.

\section{Conclusions}
\label{sec:conclusions}

In this paper we have considered the possibility of dark matter decay into Standard Model particles driven by gravitational interactions. This assumption is in line with the empirical fact that only dark matter interactions with gravity are known to exist with certainty. More precisely, we have considered operators inducing dark matter decay via nonminimal coupling to gravity for a number of common dark matter candidates, namely the scalar singlet, the inert scalar doublet and the fermionic singlet. In all these scenarios the dominant operator inducing decay is unique and proportional to the Ricci scalar $R$. 

The gravity portal induces interactions {\it i)} with the Higgs sector of the Standard Model, with strength proportional to the Standard Model masses, and {\it ii)} with the gauge-fermion sector of the Standard Model, which are mass-independent. Interactions with the gauge boson kinetic terms are absent. The dominant decay modes turn out to be those involving $W$ and $Z$ gauge bosons in the final state, when kinematically allowed, while decays into pairs of massless gauge bosons (photons and gluons) are only induced at loop level and are extremely suppressed. The decay branching ratios contain the dark matter mass as the only free parameter, thus making this framework remarkably predictive.

The strength of the gravity portal interaction can be constrained by requiring the dark matter lifetime to be at least as long as the age of the Universe, $\tau_{\DM}\gtrsim 4\times 10^{17}$~s. More stringent bounds can be derived from requiring that the cosmic gamma-ray, antimatter and neutrino fluxes induced in the decay do not exceed observations. Since the observational signals of most of the relevant channels have not been studied in detail, we have adopted a conservative and mass-independent lower limit on the total inverse decay width $\Gamma^{-1}_{\rm tot}\gtrsim 10^{24}$~s, which is in the ballpark of widths that can be probed using current neutrino observatories. 

\begin{figure}[t!]
\begin{center}
 \includegraphics[width=0.49\textwidth]{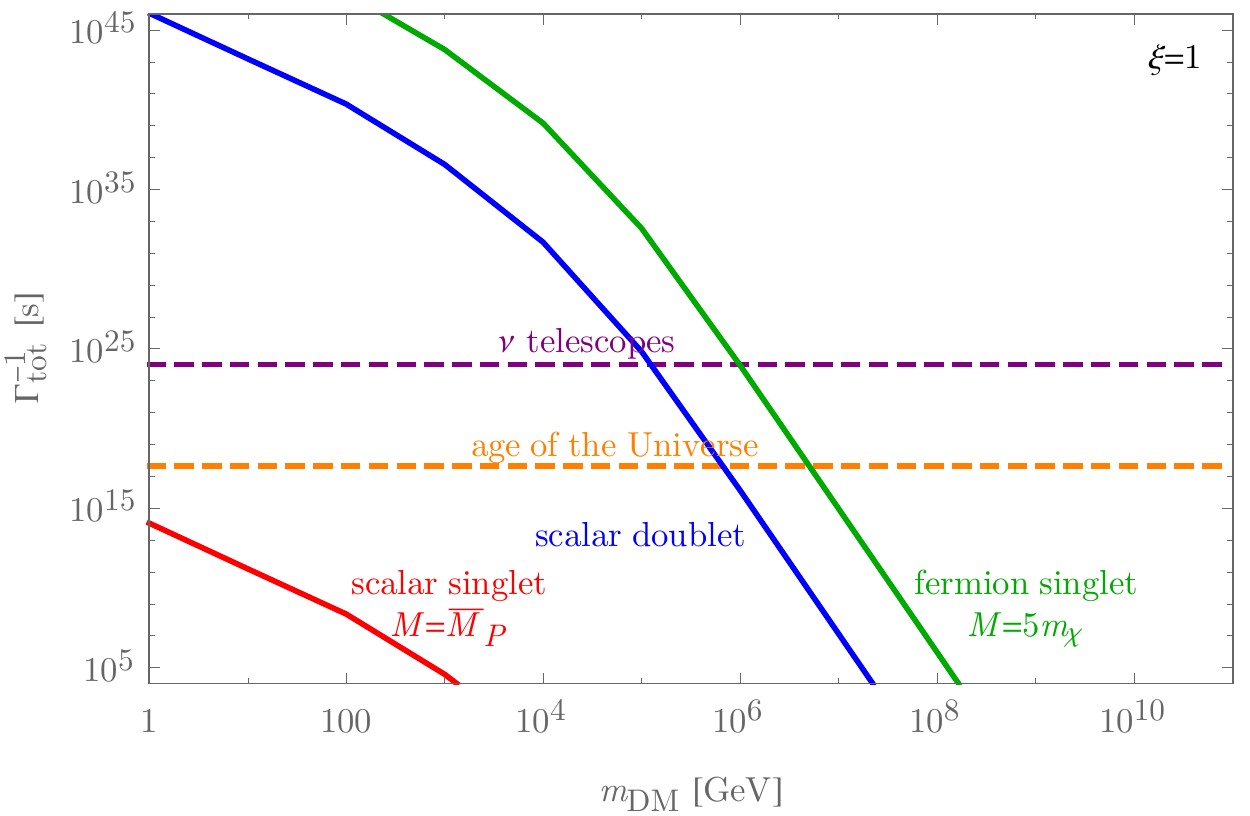} 
 \includegraphics[width=0.49\textwidth]{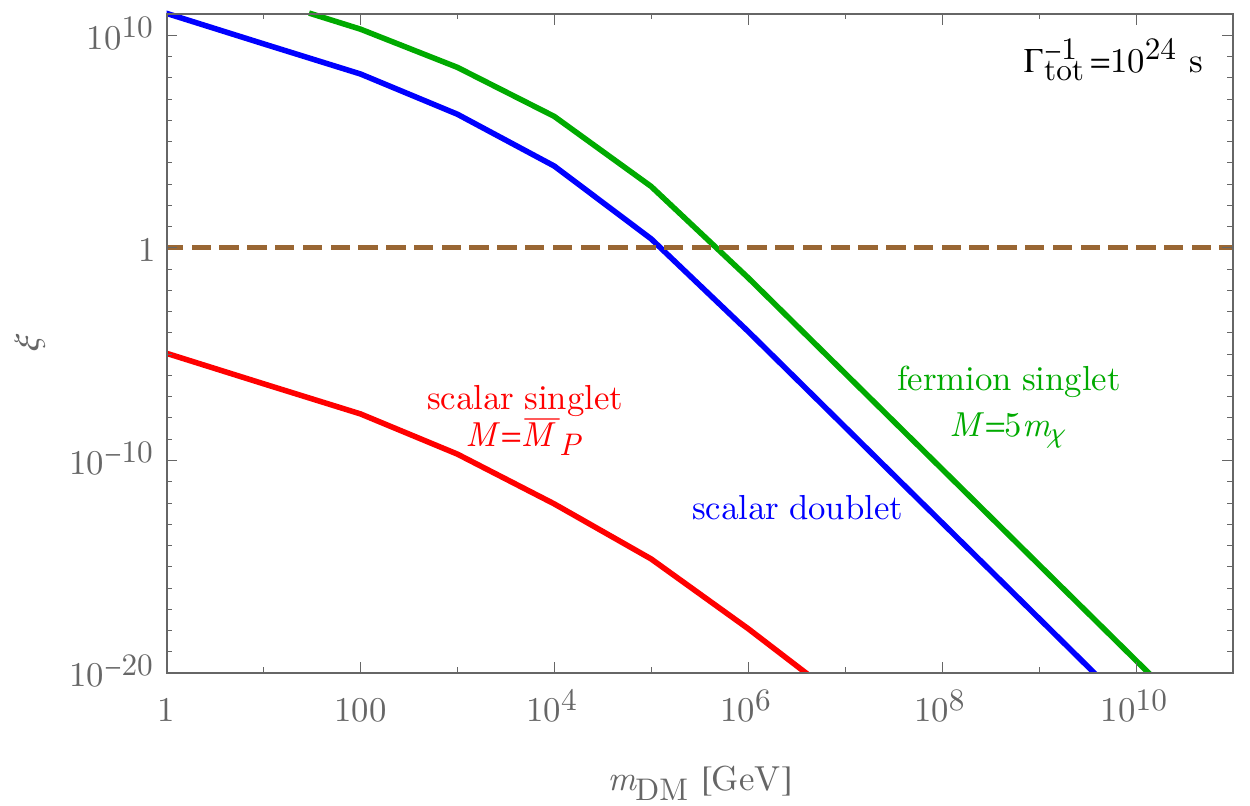} 
\end{center}
\caption{\small Total inverse widths for $\xi=1$ (left panel) and value of $\xi$ leading to $\Gamma_{\rm tot}^{-1}=10^{24}$ s (right panel) for the scalar singlet (assuming $M=\overline{M}_P$), scalar doublet and fermionic singlet (assuming $M=5m_\chi$) dark matter candidates.}
\label{fig:Comp} 
\end{figure}

Our results can be summarized in Fig.~\ref{fig:Comp}, which shows and compares the total inverse width for the three dark matter scenarios under consideration assuming $\xi=1$ (left panel), and the value of the nonminimal parameter $\xi$ which leads to a total inverse width $\Gamma^{-1}_{\rm tot}=10^{24}$ s (right panel). For the scalar singlet candidate we adopted $M=\overline{M}_P$ and for the fermionic singlet, $M=5m_\chi$. For the scalar singlet dark matter candidate, the whole range of masses studied in this paper is excluded if  $\xi\sim {\cal{O}}(1)$. Conversely, compatibility with observations requires $\xi\lesssim 10^{-9}$ for a scalar singlet mass at the electroweak scale, and $\xi\lesssim 10^{-15}$ for a mass of $\sim 100$ TeV.
The viability of the scalar singlet dark matter scenario then requires, especially for thermally produced dark matter, an additional suppression of the rate, plausibly in the form of a conserved or approximately conserved continuous symmetry. In contrast, the scalar doublet and fermionic singlet dark matter candidates are, for a wide range of masses,  naturally very long-lived, even for $\xi\sim {\cal{O}}(1)$, due to the suppression of the decay rates by factors $(v/\overline{M}_P)^2$, for the scalar doublet, and $(v/\overline{M}_P)^2 (m_{\rm DM}/M)^4$, for the fermionic singlet, where $v$ is the Higgs vacuum expectation value and $M$ is a new mass scale in the theory, related to the new physics inducing the nonminimal coupling of the dark matter to gravity. In fact, these two candidates are predicted  to be sufficiently long-lived to comply with current experimental constraints for dark matter masses below $\sim 10^5$ GeV for the scalar doublet, and $\sim 10^{6}$ GeV for the fermionic singlet, provided $\xi\lesssim 1$ and $M\gtrsim 5m_{\rm DM}$. This range of masses covers most of the window favored by dark matter production via thermal freeze-out. Therefore, for this particularly well-motivated class of candidates, provided the nonminimal coupling is sizable, signals from dark matter decay via the gravity portal could be within the reach of future indirect searches using gamma-rays, antimatter particles or neutrinos as messengers.

Our analysis has concentrated on dark matter masses above the GeV scale, where the relevant degrees of freedom are quarks, leptons, gauge bosons and the Higgs. In contrast, in the sub-GeV regime the relevant degrees of freedom are photons, leptons and hadrons. Hence, in this case, the analysis presented in this paper should be accordingly modified. This analysis will be presented in a forthcoming publication~\cite{CII:2016}.

\section*{Acknowledgements}
This work was supported in part by the DFG grant BU 1391/2-1, the DFG cluster of excellence EXC 153 ``Origin and Structure of the Universe'' and the Munich Institute for Astro- and Particle Physics (MIAPP). O.~C. is grateful to the Mainz Institute for Theoretical Physics (MITP) for its hospitality and support during the last stages of this work.

\appendix
\numberwithin{equation}{section} 
\section{Feynman rules for the decay vertices}
\label{app:rules}

\subsection{Scalar singlet}
The list of vertex rules relevant for dark matter decay follow from reading off the interactions of Eq.~(\ref{eq:LSM_ScalarDM_Einstein}). They result in 

\begin{figure}[H]
\begin{center}
\includegraphics[clip,trim={2.2cm 11.0cm 1cm 1.2cm},width=5.7in]{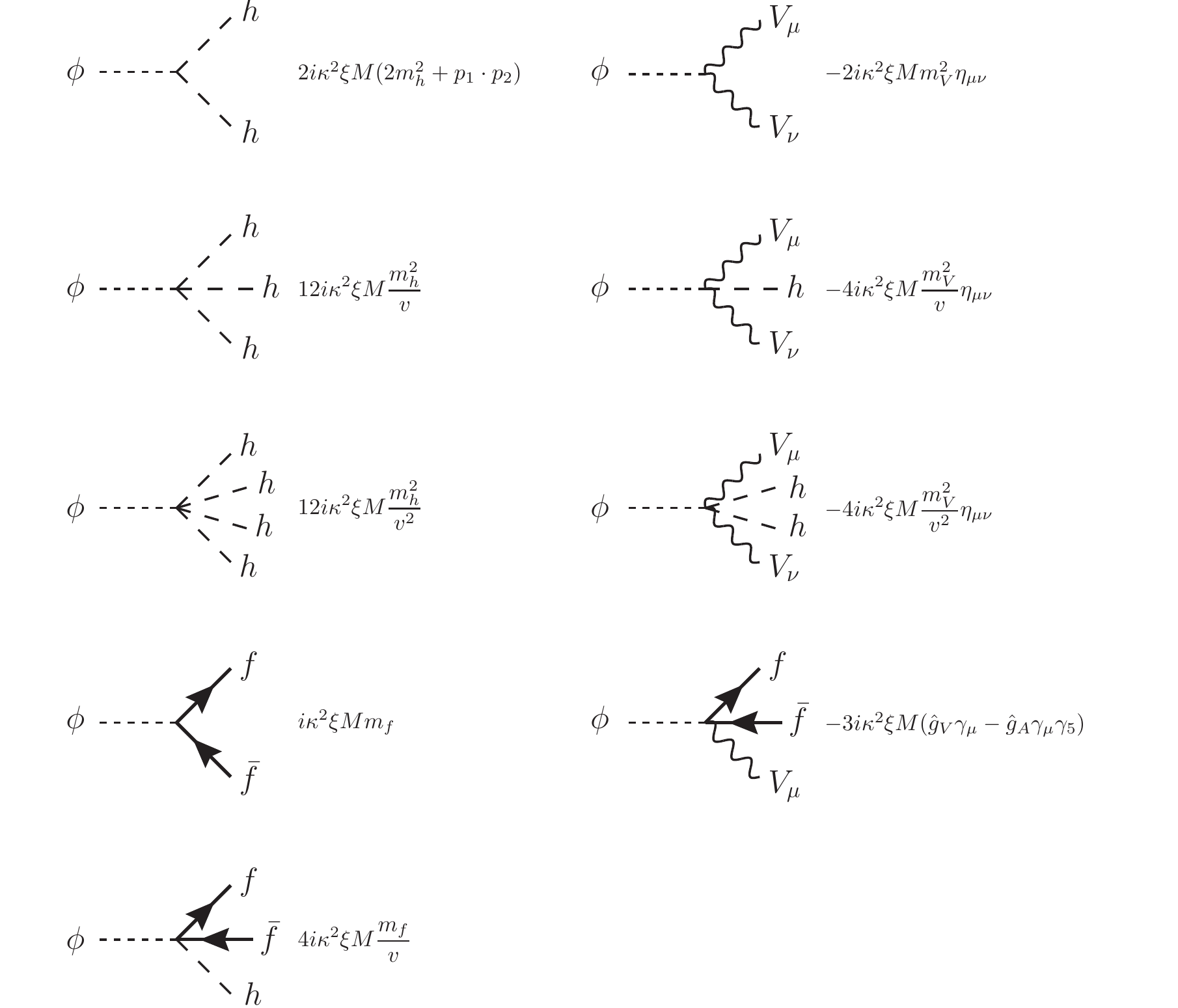}
\end{center}
\end{figure} 

The left panel shows the interactions to the pure Higgs and Yukawa sectors.
The momentum dependence in $\phi\to hh$ comes from the operator $\phi~\partial_{\mu}h\partial^{\mu}h$. This momentum dependence can be removed through integration by parts and application of the equations of motion of the leading order Standard Model Lagrangian. We have decided to keep them explicit for simplicity. Instead, the operator $\phi{\overline{f}}\,{\slash{\!\!\!\partial}}\,f$ has been simplified using the Dirac equation.

The right panel collects the gauge interactions, where $V_{\mu}=\{W_{\mu},Z_{\mu},A_{\mu},G_{\mu}\}$. The first three diagrams are proportional to the gauge boson masses and thus only exist for $V_{\mu}=W_{\mu},Z_{\mu}$. In the last diagram $\hat{g}_V$ and $\hat{g}_A$ depend on the specific gauge boson and are given by 
\begin{align}
&W_{\mu}: & \hat{g}_V&=\frac{g}{2\sqrt{2}}; & \hat{g}_A&=\frac{g}{2\sqrt{2}}\nonumber\\
&Z_{\mu}: & \hat{g}_V&=\frac{g}{2\cos{\theta_W}}(t_3-2Q\sin^2\theta_W); & \hat{g}_A&=\frac{g}{2\cos{\theta_W}}t_3\nonumber\\
&A_{\mu}: & \hat{g}_V&=Qe; & \hat{g}_A&=0 \nonumber\\
&G_{\mu}: & \hat{g}_V&=g_s t^a; & \hat{g}_A&=0
\end{align}
\subsection{Inert doublet}
The interactions can be read off from Eq.~(\ref{eq:LSM_Doublet_DM_Einstein}) and are given by

\begin{figure}[H]
\begin{center}
\includegraphics[clip,trim={2.2cm 4.2cm 1.2cm 1.1cm},width=5.5in]{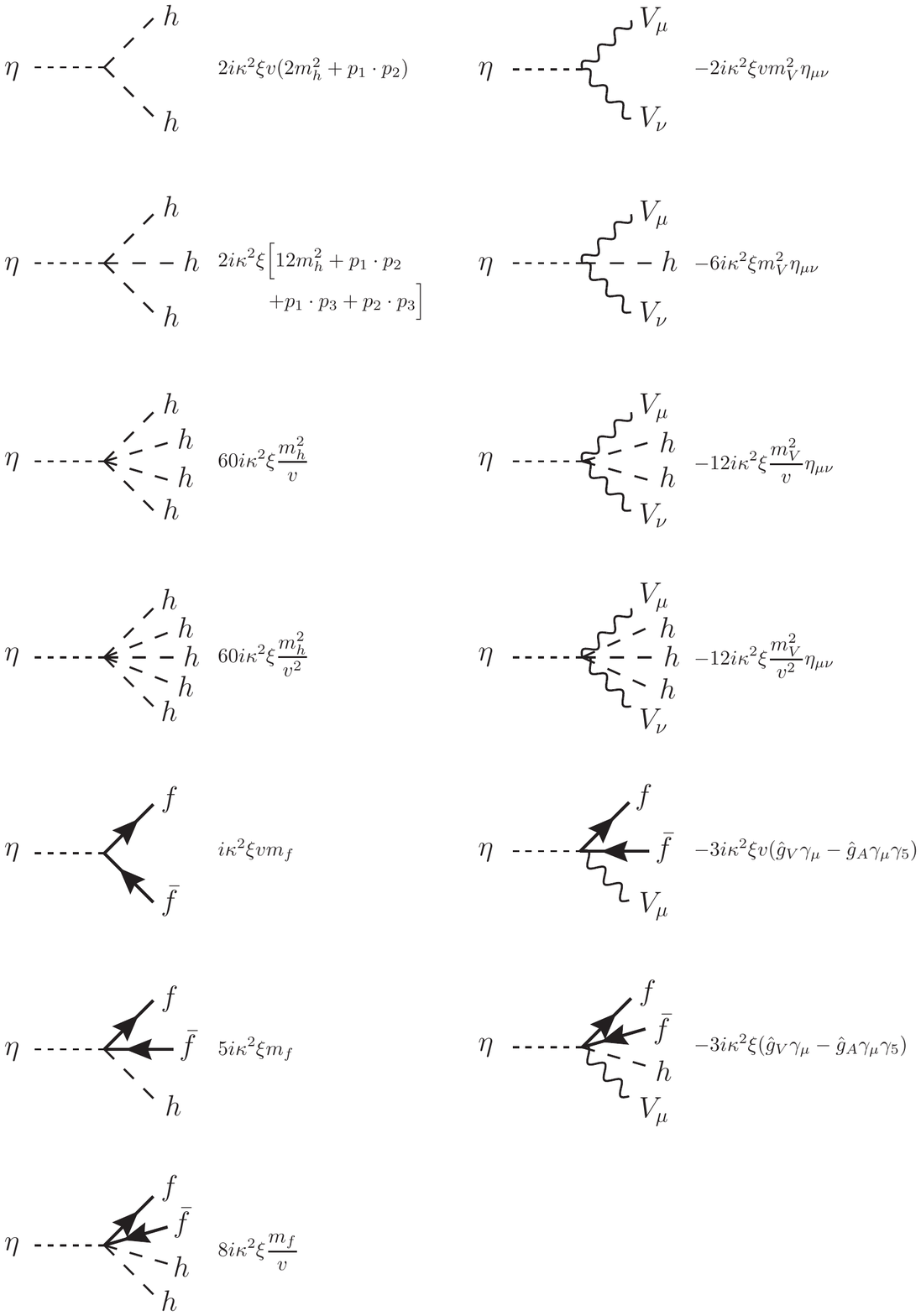}
\end{center}
\end{figure} 

The same conventions as in the scalar singlet case apply.
\subsection{Fermionic singlet}
The interactions can be read off from Eq.~(\ref{eq:fermionFR}). The resulting vertex topologies are the same as in the inert doublet case, replacing $\eta\to \chi$ and appending a neutrino leg to each vertex. The Feynman rules are then determined by adding the fermionic current 
\begin{align}
\frac{1}{\sqrt{2}M^2}\overline{u}(p_{\nu})\frac{1-\gamma_5}{2}u(p_{\chi})\;,
\end{align}
to the list of decay vertices of the inert doublet case.

\section{Kinematics of $n$-body decays}
\label{app:kinematics}
The decay rate for an arbitrary $n$-body decay is given by the master formula:
\begin{align}
d\Gamma_{\varphi\to n}&=\frac{1}{2M_{\varphi}}|{\cal{M}}|^2 d\Phi_n\;,
\end{align}
where $\varphi$ is a generic dark matter candidate (for the purposes of this paper, a fermionic singlet, a scalar singlet or a scalar doublet), $d\Phi_n$ is the differential $n$-body phase space 
\begin{align}
d\Phi_n&=(2\pi)^4\delta^4\bigg(P_{\varphi}-\sum_j^np_j\bigg)\prod_j^n\frac{d^3p_j}{(2\pi)^3 2E_j}\;,
\end{align}
and $|{\cal{M}}|^2$ corresponds to the unpolarized $n$-body decay amplitude squared, i.e., summed over final-state polarizations and averaged over initial spin. Symmetry factors are implicitly included.

In this paper we needed to consider up to six-body decays. Given the technical complexity involved in the evaluation of the phase space, for computational purposes it is convenient to reduce the problem into sequential chains of two-body decays, computed in the respective center-of-mass frames, and eventually boosted to the rest frame of the dark matter particle. For a generic $n$-body decay, this can be done by introducing a set of $(n-2)$ intermediate virtual states integrated over the full virtuality range. Cast in this form, the $(3n-4)$ integrals to be performed get split into $(2n-2)$ angular integrals, defined in the different two-body CM frames, and $(n-2)$ integrals over the intermediate virtual state energies, all of them with rather trivial integration limits.

As an illustrative example, consider three-body decays with the final-state momenta $p_1$ to $p_3$. We split the original decay $P_{\varphi}\to p_1p_2p_3$ as the sequential $P\to s_{12} p_3\to p_1p_2p_3$, with the angles defined in the CM frame of the two subsystems, CM$_{12}$ and CM$_3$. With this splitting choice, the phase space can be written as
\begin{align}
d\Phi_3&=\frac{ds_{12}}{2\pi}d\Phi_2(s_{12},p_3)d\Phi_2(p_1,p_2)\;,
\end{align}
with basic building blocks  
\begin{align}
d\Phi_2(p_1,p_2)&=\frac{\lambda_{12}}{8(2\pi)^2}d\Omega_{12},\quad \lambda_{12}(m_1^2,m_2^2,s_{12})=\left[1-2\frac{(m_1^2+m_2^2)}{s_{12}}+\frac{(m_1^2-m_2^2)^2}{s_{12}^2}\right]^{1/2}\;,\\
d\Phi_2(s_{12},p_3)&=\frac{\lambda_0}{8(2\pi)^2}d\Omega_{3},\quad \lambda_0(s_{12},m_3^2,M_{\varphi}^2)=\left[1-2\frac{(s_{12}+m_3^2)}{M_{\varphi}^2}+\frac{(s_{12}-m_3^2)^2}{M_{\varphi}^4}\right]^{1/2}\;,
\end{align}
and $(m_1+m_2)^2\leq s_{12}\leq (M_{\varphi}-m_3)^2$. The kinematic invariants in the $\varphi$ CM frame in terms of the CM$_{12}$ and CM$_3$ angles are
\begin{align}
p_1\cdot p_2&=\frac{s_{12}-m_1^2-m_2^2}{2}\;,\nonumber\\
p_1\cdot p_3&=\frac{1}{4}\left[\frac{(M_{\varphi}^2-s_{12}-m_3^2)(s_{12}+m_1^2-m_2^2)}{s_{12}}+M_{\varphi}^2\lambda_0\lambda_{12}c_{\alpha}\right]\;,\nonumber\\
p_2\cdot p_3&=\frac{1}{4}\left[\frac{(M_{\varphi}^2-s_{12}-m_3^2)(s_{12}+m_2^2-m_1^2)}{s_{12}}-M_{\varphi}^2\lambda_0\lambda_{12}c_{\alpha}\right]\;.
\end{align}
They can be easily constructed by defining the momenta in the CM$_{12}$ and CM$_3$ frames, which point in the directions
\begin{align}
{\hat{n}}_1=(c_{\alpha},s_{\alpha},0);\quad {\hat{n}}_2=(c_{\alpha},-s_{\alpha},0);\quad {\hat{n}}_3=(0,0,0)\;,
\end{align}
and boosting them to the $\varphi$ CM frame, which can be achieved with velocities
\begin{align}
\beta_{12}=\frac{\lambda_0 M_{\varphi}^2}{M_{\varphi}^2+s_{12}-m_3^2};\qquad \beta_3=\frac{\lambda_0 M_{\varphi}^2}{M_{\varphi}^2+m_3^2-s_{12}}\;. 
\end{align}
The full differential decay rate can be computed by convoluting the phase space with the corresponding $|{\cal{M}}|^2$, whose explicit expressions can be computed from the Feynman rules listed in Appendix A. Using the formulas above, $|{\cal{M}}|^2$ can be brought to be a function only of $s_{12}$ and CM angles, which can be numerically integrated. 

By iterating the method, one can find expressions for general $n$-body decays. For instance, defining $\Sigma_{ij\dots k}\equiv (m_i+m_j+\dots m_k)$, the four-, five- and six-body phase space formulae can be written as
\begin{align}
d\Phi_4&=\frac{ds_{12}}{2\pi}\frac{ds_{34}}{2\pi}d\Phi_2(s_{12},s_{34})d\Phi_2(p_1,p_2)d\Phi_2(p_3,p_4),
\end{align}
with $\Sigma_{12}\leq \sqrt{s_{12}}\leq M_{\varphi}-\Sigma_{34}$ and $\Sigma_{34}\leq \sqrt{s_{34}}\leq M_{\varphi}-\sqrt{s_{12}}$;
\begin{align}
d\Phi_5&=\frac{ds_{12}}{2\pi}\frac{ds_{345}}{2\pi}\frac{ds_{45}}{2\pi}d\Phi_2(s_{12},s_{345})d\Phi_2(p_3,s_{45})d\Phi_2(p_1,p_2)d\Phi_2(p_4,p_5),
\end{align}
with $\Sigma_{12}\leq \sqrt{s_{12}}\leq M_{\varphi}-\Sigma_{345}$, $\Sigma_{345}\leq \sqrt{s_{345}}\leq M_{\varphi}-\sqrt{s_{12}}$ and $\Sigma_{45}\leq \sqrt{s_{45}}\leq \sqrt{s_{345}}-m_3$; and
\begin{align}
d\Phi_6&=\frac{ds_{12}}{2\pi}\frac{ds_{3456}}{2\pi}\frac{ds_{456}}{2\pi}\frac{ds_{56}}{2\pi}d\Phi_2(s_{12},s_{3456})d\Phi_2(p_3,s_{456})d\Phi_2(p_4,s_{56})d\Phi_2(p_1,p_2)d\Phi_2(p_5,p_6),
\end{align}
with $\Sigma_{12}\leq \sqrt{s_{12}}\leq M_{\varphi}-\Sigma_{3456}$, $\Sigma_{3456}\leq \sqrt{s_{3456}}\leq M_{\varphi}-\sqrt{s_{12}}$, $\Sigma_{456}\leq \sqrt{s_{456}}\leq \sqrt{s_{3456}}-m_3$ and $\Sigma_{56}\leq \sqrt{s_{56}}\leq \sqrt{s_{456}}-m_4$.
The final form of the differential decay rate will be a function of the $n!/(2!(n-2)!)$ independent kinematic invariants, written in terms of CM angles and invariant masses.

The results above obviously do not depend on the chosen two-body splitting configuration. However, a wise choice of variables can save considerable computational effort. In general, the simplest configuration is dictated by the dynamics of the system, encoded in $|{\cal{M}}|^2$. In particular, the splitting configuration chosen above is adapted to decays of the form $\varphi\to WW+ nh$. For this family of decays, the interaction amplitude only depends on the invariant $p_1\cdot p_2=\tfrac{1}{2}(s_{12}-m_1^2-m_2^2)$ and therefore the integrals get simplified.

\bibliography{references}

\providecommand{\href}[2]{#2}\begingroup\raggedright\begin{thebibliography}{10}

\bibitem{Bertone:2010zza}
G.~Bertone, ed., \emph{{Particle Dark Matter: Observations, Models and
  Searches}}.
\newblock {Cambridge U. Press}, 2010.

\bibitem{Bergstrom:2000pn}
L.~Bergstrom, \emph{{Nonbaryonic dark matter: Observational evidence and
  detection methods}},
  \href{http://dx.doi.org/10.1088/0034-4885/63/5/2r3}{\emph{Rept. Prog. Phys.}
  {\bf 63} (2000) 793}, [\href{https://arxiv.org/abs/hep-ph/0002126}{{\tt
  hep-ph/0002126}}].

\bibitem{Bertone:2004pz}
G.~Bertone, D.~Hooper and J.~Silk, \emph{{Particle dark matter: Evidence,
  candidates and constraints}},
  \href{http://dx.doi.org/10.1016/j.physrep.2004.08.031}{\emph{Phys. Rept.}
  {\bf 405} (2005) 279--390}, [\href{https://arxiv.org/abs/hep-ph/0404175}{{\tt
  hep-ph/0404175}}].

\bibitem{Ibarra:2013cra}
A.~Ibarra, D.~Tran and C.~Weniger, \emph{{Indirect Searches for Decaying Dark
  Matter}}, \href{http://dx.doi.org/10.1142/S0217751X13300408}{\emph{Int. J.
  Mod. Phys.} {\bf A28} (2013) 1330040},
  [\href{https://arxiv.org/abs/1307.6434}{{\tt 1307.6434}}].

\bibitem{Abbott:1982af}
L.~F. Abbott and P.~Sikivie, \emph{{A Cosmological Bound on the Invisible
  Axion}}, \href{http://dx.doi.org/10.1016/0370-2693(83)90638-X}{\emph{Phys.
  Lett.} {\bf B120} (1983) 133--136}.

\bibitem{Preskill:1982cy}
J.~Preskill, M.~B. Wise and F.~Wilczek, \emph{{Cosmology of the Invisible
  Axion}}, \href{http://dx.doi.org/10.1016/0370-2693(83)90637-8}{\emph{Phys.
  Lett.} {\bf B120} (1983) 127--132}.

\bibitem{Dine:1982ah}
M.~Dine and W.~Fischler, \emph{{The Not So Harmless Axion}},
  \href{http://dx.doi.org/10.1016/0370-2693(83)90639-1}{\emph{Phys. Lett.} {\bf
  B120} (1983) 137--141}.

\bibitem{Dodelson:1993je}
S.~Dodelson and L.~M. Widrow, \emph{{Sterile-neutrinos as dark matter}},
  \href{http://dx.doi.org/10.1103/PhysRevLett.72.17}{\emph{Phys. Rev. Lett.}
  {\bf 72} (1994) 17--20}, [\href{https://arxiv.org/abs/hep-ph/9303287}{{\tt
  hep-ph/9303287}}].

\bibitem{Shi:1998km}
X.-D. Shi and G.~M. Fuller, \emph{{A New dark matter candidate: Nonthermal
  sterile neutrinos}},
  \href{http://dx.doi.org/10.1103/PhysRevLett.82.2832}{\emph{Phys. Rev. Lett.}
  {\bf 82} (1999) 2832--2835},
  [\href{https://arxiv.org/abs/astro-ph/9810076}{{\tt astro-ph/9810076}}].

\bibitem{Abazajian:2001nj}
K.~Abazajian, G.~M. Fuller and M.~Patel, \emph{{Sterile neutrino hot, warm, and
  cold dark matter}},
  \href{http://dx.doi.org/10.1103/PhysRevD.64.023501}{\emph{Phys. Rev.} {\bf
  D64} (2001) 023501}, [\href{https://arxiv.org/abs/astro-ph/0101524}{{\tt
  astro-ph/0101524}}].

\bibitem{Boehm:2003hm}
C.~Boehm and P.~Fayet, \emph{{Scalar dark matter candidates}},
  \href{http://dx.doi.org/10.1016/j.nuclphysb.2004.01.015}{\emph{Nucl. Phys.}
  {\bf B683} (2004) 219--263},
  [\href{https://arxiv.org/abs/hep-ph/0305261}{{\tt hep-ph/0305261}}].

\bibitem{Griest:1989wd}
K.~Griest and M.~Kamionkowski, \emph{{Unitarity Limits on the Mass and Radius
  of Dark Matter Particles}},
  \href{http://dx.doi.org/10.1103/PhysRevLett.64.615}{\emph{Phys. Rev. Lett.}
  {\bf 64} (1990) 615}.

\bibitem{Chung:1998ua}
D.~J.~H. Chung, E.~W. Kolb and A.~Riotto, \emph{{Nonthermal supermassive dark
  matter}}, \href{http://dx.doi.org/10.1103/PhysRevLett.81.4048}{\emph{Phys.
  Rev. Lett.} {\bf 81} (1998) 4048--4051},
  [\href{https://arxiv.org/abs/hep-ph/9805473}{{\tt hep-ph/9805473}}].

\bibitem{Cata:2016dsg}
O.~Cat{\`a}, A.~Ibarra and S.~Ingenh{\"u}tt, \emph{{Dark matter decays from
  nonminimal coupling to gravity}},
  \href{http://dx.doi.org/10.1103/PhysRevLett.117.021302}{\emph{Phys. Rev.
  Lett.} {\bf 117} (2016) 021302},
  [\href{https://arxiv.org/abs/1603.03696}{{\tt 1603.03696}}].

\bibitem{Salopek:1988qh}
D.~S. Salopek, J.~R. Bond and J.~M. Bardeen, \emph{{Designing Density
  Fluctuation Spectra in Inflation}},
  \href{http://dx.doi.org/10.1103/PhysRevD.40.1753}{\emph{Phys. Rev.} {\bf D40}
  (1989) 1753}.

\bibitem{Ren:2014sya}
J.~Ren, Z.-Z. Xianyu and H.-J. He, \emph{{Higgs Gravitational Interaction, Weak
  Boson Scattering, and Higgs Inflation in Jordan and Einstein Frames}},
  \href{http://dx.doi.org/10.1088/1475-7516/2014/06/032}{\emph{JCAP} {\bf 1406}
  (2014) 032}, [\href{https://arxiv.org/abs/1404.4627}{{\tt 1404.4627}}].

\bibitem{Ren:2014mta}
J.~Ren and H.-J. He, \emph{{Probing Gravitational Dark Matter}},
  \href{http://dx.doi.org/10.1088/1475-7516/2015/03/052}{\emph{JCAP} {\bf 1503}
  (2015) 052}, [\href{https://arxiv.org/abs/1410.6436}{{\tt 1410.6436}}].

\bibitem{Tang:2016vch}
Y.~Tang and Y.-L. Wu, \emph{{Pure Gravitational Dark Matter, Its Mass and
  Signatures}},
  \href{http://dx.doi.org/10.1016/j.physletb.2016.05.045}{\emph{Phys. Lett.}
  {\bf B758} (2016) 402--406}, [\href{https://arxiv.org/abs/1604.04701}{{\tt
  1604.04701}}].

\bibitem{Mambrini:2015sia}
Y.~Mambrini, S.~Profumo and F.~S. Queiroz, \emph{{Dark Matter and Global
  Symmetries}},
  \href{http://dx.doi.org/10.1016/j.physletb.2016.07.076}{\emph{Phys. Lett.}
  {\bf B760} (2016) 807--815}, [\href{https://arxiv.org/abs/1508.06635}{{\tt
  1508.06635}}].

\bibitem{CII:2016}
O.~Cat{\`a}, A.~Ibarra and S.~Ingenh{\"u}tt.
\newblock {In preparation.}

\bibitem{Cirelli:2012ut}
M.~Cirelli, E.~Moulin, P.~Panci, P.~D. Serpico and A.~Viana, \emph{{Gamma ray
  constraints on Decaying Dark Matter}},
  \href{http://dx.doi.org/10.1103/PhysRevD.86.083506,
  10.1103/PhysRevD.86.109901}{\emph{Phys. Rev.} {\bf D86} (2012) 083506},
  [\href{https://arxiv.org/abs/1205.5283}{{\tt 1205.5283}}].

\bibitem{Ibarra:2013zia}
A.~Ibarra, A.~S. Lamperstorfer and J.~Silk, \emph{{Dark matter annihilations
  and decays after the AMS-02 positron measurements}},
  \href{http://dx.doi.org/10.1103/PhysRevD.89.063539}{\emph{Phys. Rev.} {\bf
  D89} (2014) 063539}, [\href{https://arxiv.org/abs/1309.2570}{{\tt
  1309.2570}}].

\bibitem{Giesen:2015ufa}
G.~Giesen, M.~Boudaud, Y.~G{\'e}nolini, V.~Poulin, M.~Cirelli, P.~Salati
  et~al., \emph{{AMS-02 antiprotons, at last! Secondary astrophysical component
  and immediate implications for Dark Matter}},
  \href{http://dx.doi.org/10.1088/1475-7516/2015/09/023,
  10.1088/1475-7516/2015/9/023}{\emph{JCAP} {\bf 1509} (2015) 023},
  [\href{https://arxiv.org/abs/1504.04276}{{\tt 1504.04276}}].

\bibitem{Accardo:2014lma}
{\scshape AMS} collaboration, L.~Accardo et~al., \emph{{High Statistics
  Measurement of the Positron Fraction in Primary Cosmic Rays of 0.5–500 GeV
  with the Alpha Magnetic Spectrometer on the International Space Station}},
  \href{http://dx.doi.org/10.1103/PhysRevLett.113.121101}{\emph{Phys. Rev.
  Lett.} {\bf 113} (2014) 121101}.

\bibitem{Aguilar:2016kjl}
{\scshape AMS} collaboration, M.~Aguilar et~al., \emph{{Antiproton Flux,
  Antiproton-to-Proton Flux Ratio, and Properties of Elementary Particle Fluxes
  in Primary Cosmic Rays Measured with the Alpha Magnetic Spectrometer on the
  International Space Station}},
  \href{http://dx.doi.org/10.1103/PhysRevLett.117.091103}{\emph{Phys. Rev.
  Lett.} {\bf 117} (2016) 091103}.

\bibitem{Esmaili:2012us}
A.~Esmaili, A.~Ibarra and O.~L.~G. Peres, \emph{{Probing the stability of
  superheavy dark matter particles with high-energy neutrinos}},
  \href{http://dx.doi.org/10.1088/1475-7516/2012/11/034}{\emph{JCAP} {\bf 1211}
  (2012) 034}, [\href{https://arxiv.org/abs/1205.5281}{{\tt 1205.5281}}].

\bibitem{Silveira:1985rk}
V.~Silveira and A.~Zee, \emph{{Scalar phantoms}},
  \href{http://dx.doi.org/10.1016/0370-2693(85)90624-0}{\emph{Phys. Lett.} {\bf
  B161} (1985) 136}.

\bibitem{McDonald:1993ex}
J.~McDonald, \emph{{Gauge singlet scalars as cold dark matter}},
  \href{http://dx.doi.org/10.1103/PhysRevD.50.3637}{\emph{Phys. Rev.} {\bf D50}
  (1994) 3637--3649}, [\href{https://arxiv.org/abs/hep-ph/0702143}{{\tt
  hep-ph/0702143}}].

\bibitem{Cline:2013gha}
J.~M. Cline, K.~Kainulainen, P.~Scott and C.~Weniger, \emph{{Update on scalar
  singlet dark matter}}, \href{http://dx.doi.org/10.1103/PhysRevD.92.039906,
  10.1103/PhysRevD.88.055025}{\emph{Phys. Rev.} {\bf D88} (2013) 055025},
  [\href{https://arxiv.org/abs/1306.4710}{{\tt 1306.4710}}].

\bibitem{Alwall:2014hca}
J.~Alwall, R.~Frederix, S.~Frixione, V.~Hirschi, F.~Maltoni, O.~Mattelaer
  et~al., \emph{{The automated computation of tree-level and next-to-leading
  order differential cross sections, and their matching to parton shower
  simulations}}, \href{http://dx.doi.org/10.1007/JHEP07(2014)079}{\emph{JHEP}
  {\bf 07} (2014) 079}, [\href{https://arxiv.org/abs/1405.0301}{{\tt
  1405.0301}}].

\bibitem{Deshpande:1977rw}
N.~G. Deshpande and E.~Ma, \emph{{Pattern of Symmetry Breaking with Two Higgs
  Doublets}},
  \href{http://dx.doi.org/10.1103/PhysRevD.18.2574}{\emph{Phys.Rev.} {\bf D18}
  (1978) 2574}.

\bibitem{LopezHonorez:2006gr}
L.~Lopez~Honorez, E.~Nezri, J.~F. Oliver and M.~H.~G. Tytgat, \emph{{The Inert
  Doublet Model: An Archetype for Dark Matter}},
  \href{http://dx.doi.org/10.1088/1475-7516/2007/02/028}{\emph{JCAP} {\bf 0702}
  (2007) 028}, [\href{https://arxiv.org/abs/hep-ph/0612275}{{\tt
  hep-ph/0612275}}].

\bibitem{Aad:2012tfa}
{\scshape ATLAS} collaboration, G.~Aad et~al., \emph{{Observation of a new
  particle in the search for the Standard Model Higgs boson with the ATLAS
  detector at the LHC}},
  \href{http://dx.doi.org/10.1016/j.physletb.2012.08.020}{\emph{Phys. Lett.}
  {\bf B716} (2012) 1--29}, [\href{https://arxiv.org/abs/1207.7214}{{\tt
  1207.7214}}].

\bibitem{Chatrchyan:2012xdj}
{\scshape CMS} collaboration, S.~Chatrchyan et~al., \emph{{Observation of a new
  boson at a mass of 125 GeV with the CMS experiment at the LHC}},
  \href{http://dx.doi.org/10.1016/j.physletb.2012.08.021}{\emph{Phys. Lett.}
  {\bf B716} (2012) 30--61}, [\href{https://arxiv.org/abs/1207.7235}{{\tt
  1207.7235}}].

\bibitem{Aartsen:2013bka}
{\scshape IceCube} collaboration, M.~G. Aartsen et~al., \emph{{First
  observation of PeV-energy neutrinos with IceCube}},
  \href{http://dx.doi.org/10.1103/PhysRevLett.111.021103}{\emph{Phys. Rev.
  Lett.} {\bf 111} (2013) 021103}, [\href{https://arxiv.org/abs/1304.5356}{{\tt
  1304.5356}}].

\bibitem{Aartsen:2014gkd}
{\scshape IceCube} collaboration, M.~G. Aartsen et~al., \emph{{Observation of
  High-Energy Astrophysical Neutrinos in Three Years of IceCube Data}},
  \href{http://dx.doi.org/10.1103/PhysRevLett.113.101101}{\emph{Phys. Rev.
  Lett.} {\bf 113} (2014) 101101}, [\href{https://arxiv.org/abs/1405.5303}{{\tt
  1405.5303}}].

\bibitem{Feldstein:2013kka}
B.~Feldstein, A.~Kusenko, S.~Matsumoto and T.~T. Yanagida, \emph{{Neutrinos at
  IceCube from Heavy Decaying Dark Matter}},
  \href{http://dx.doi.org/10.1103/PhysRevD.88.015004}{\emph{Phys. Rev.} {\bf
  D88} (2013) 015004}, [\href{https://arxiv.org/abs/1303.7320}{{\tt
  1303.7320}}].

\bibitem{Esmaili:2013gha}
A.~Esmaili and P.~D. Serpico, \emph{{Are IceCube neutrinos unveiling PeV-scale
  decaying dark matter?}},
  \href{http://dx.doi.org/10.1088/1475-7516/2013/11/054}{\emph{JCAP} {\bf 1311}
  (2013) 054}, [\href{https://arxiv.org/abs/1308.1105}{{\tt 1308.1105}}].

\bibitem{Kim:2006af}
Y.~G. Kim and K.~Y. Lee, \emph{{The Minimal model of fermionic dark matter}},
  \href{http://dx.doi.org/10.1103/PhysRevD.75.115012}{\emph{Phys. Rev.} {\bf
  D75} (2007) 115012}, [\href{https://arxiv.org/abs/hep-ph/0611069}{{\tt
  hep-ph/0611069}}].

\bibitem{Garny:2015wea}
M.~Garny, A.~Ibarra and S.~Vogl, \emph{{Signatures of Majorana dark matter with
  t-channel mediators}},
  \href{http://dx.doi.org/10.1142/S0218271815300190}{\emph{Int. J. Mod. Phys.}
  {\bf D24} (2015) 1530019}, [\href{https://arxiv.org/abs/1503.01500}{{\tt
  1503.01500}}].

\bibitem{Griest:1990kh}
K.~Griest and D.~Seckel, \emph{{Three exceptions in the calculation of relic
  abundances}}, \href{http://dx.doi.org/10.1103/PhysRevD.43.3191}{\emph{Phys.
  Rev.} {\bf D43} (1991) 3191--3203}.

\bibitem{Flores:1989ru}
R.~Flores, K.~A. Olive and S.~Rudaz, \emph{{Radiative Processes in Lsp
  Annihilation}},
  \href{http://dx.doi.org/10.1016/0370-2693(89)90760-0}{\emph{Phys. Lett.} {\bf
  B232} (1989) 377--382}.

\bibitem{Bergstrom:1989jr}
L.~Bergstrom, \emph{{Radiative Processes in Dark Matter Photino Annihilation}},
  \href{http://dx.doi.org/10.1016/0370-2693(89)90585-6}{\emph{Phys. Lett.} {\bf
  B225} (1989) 372}.

\bibitem{Bringmann:2007nk}
T.~Bringmann, L.~Bergstrom and J.~Edsjo, \emph{{New Gamma-Ray Contributions to
  Supersymmetric Dark Matter Annihilation}},
  \href{http://dx.doi.org/10.1088/1126-6708/2008/01/049}{\emph{JHEP} {\bf 01}
  (2008) 049}, [\href{https://arxiv.org/abs/0710.3169}{{\tt 0710.3169}}].

\end{thebibliography}\endgroup

\end{document}